\documentclass[fleqn,usenatbib]{mnras}

\usepackage{newtxtext,newtxmath}

\usepackage[T1]{fontenc}

\DeclareRobustCommand{\VAN}[3]{#2}
\let\VANthebibliography\thebibliography
\def\thebibliography{\DeclareRobustCommand{\VAN}[3]{##3}\VANthebibliography}

\usepackage{graphicx}	
\usepackage{amsmath}	
\usepackage{aasmacros}

\DeclareMathOperator*{\argmin}{argmin}
\title[Measuring atmospheric dispersion]{Demonstration of an imaging technique for the measurement of PSF elongation caused by Atmospheric Dispersion}

\author[J.A. van den Born et al.]{ 
J.A. van den Born,$^{1,2,3}$\thanks{E-mail: born@astro.rug.nl} 
W. Jellema,$^{2,4}$
and E. Dijkstra$^{1}$
\\
$^{1}$NOVA Optical Infrared Instrumentation Group at ASTRON, Oude Hoogeveensedijk 4, NL-7991 PD Dwingeloo, the Netherlands\\
$^{2}$Kapteyn Astronomical Institute, University of Groningen, PO Box 800, NL-9700 AV Groningen, the Netherlands\\
$^{3}$Engineering and Technology Institute Groningen, University of Groningen, Nijenborgh 4, NL-9747 AG Groningen, the Netherlands\\
$^{4}$SRON Netherlands Insitute for Space Research, PO Box 800, NL-9700 AV Groningen, the Netherlands
}

\date{Accepted XXX. Received YYY; in original form ZZZ}

\pubyear{2022}

\begin{document}
\label{firstpage}
\pagerange{\pageref{firstpage}--\pageref{lastpage}}
\maketitle

\begin{abstract}
Elongation of the point spread function due to atmospheric dispersion becomes a severe problem for high resolution imaging instruments, if an atmospheric dispersion corrector is not present. In this work we report on a novel technique to measure this elongation, corrected or uncorrected, from imaging data. By employing a simple diffraction mask it is possible to magnify the chromatic elongation caused by the atmosphere and thus make it easier to measure. We discuss the theory and design of such a mask and report on two proof of concept observations using the 40 cm Gratama telescope at the University of Groningen. We evaluate the acquired images using a geometric approach, a forward modelling approach and from a direct measurement of the length of the point spread function. For the first two methods we report measurements consistent with atmospheric dispersion models to within 0.5~arcsec. Direct measurements of the elongation do not prove suitable for the characterisation of atmospheric dispersion. We conclude that the addition of this type of diffraction mask can be valuable for measurements of PSF elongation. This can enable high precision correction of atmospheric dispersion on future instruments.
\end{abstract}




\section{Introduction}
Atmospheric dispersion is the differential refraction of polychromatic light as it travels through the atmosphere. To a first approximation, it can be expressed in arcseconds as
\begin{equation}
    \Delta R = 206265 \times \left[ n(\lambda_1) - n(\lambda_2) \right] \tan z, \label{eq:atmospheric_dispersion}
\end{equation}
where $z$ is the observed zenith angle and $n(\lambda)$ is the refractive index of air at a wavelength $\lambda$.
Atmospheric dispersion presents itself as an increasing elongation of the point spread function (PSF) as the zenith distance of the observed source increases. 

Atmospheric Dispersion Correctors (ADCs) can be employed to counteract this form of image degradation (see for example \cite{Avila1997,Egner2010,terHorst2016}). While many telescopes in use today already incorporate ADCs, the increase in resolution promised by upcoming extremely large telescopes (ELTs) will make the use of an ADC an absolute necessity, assuming that diffraction limited performance is desired. The near-infrared imaging instruments for the ELTs will have to reduce the PSF elongation to less than a few milli arcseconds \citep{Phillips2016, vandenBorn2020}. 

A direct analysis of the elongation of the point spread function can sometimes be used to detect the presence of dispersion. This might then be used to verify and update the ADC configuration \citep{Cabral2020}. However, it is difficult to disentangle chromatic dispersion from other systematic effects that change the PSF shape, such as optical aberrations, telescope guiding errors, instrument flexure, and telescope vibrations. The successful application of such a direct method thus requires a complete understanding of the telescope. In practice, this will not be feasible at milli arcsecond levels, but only when the instrument requirements are relatively relaxed.

Using a different method, dispersion correction at milli arcsecond level has already been demonstrated with the ADC on Subaru's AO188 instrument in \cite{Pathak2018}. They used the projection of a waffle pattern on a deformable mirror to modify the PSF in such a way that highly accurate dispersion correction could be performed under closed-loop control. 
However, the extended shape of the PSF that results from this approach is not acceptable in many scientific cases, such as observations of crowded stellar regions or extended sources. A feed-forward control approach that does not modify the PSF will generally be preferred, even though the dispersion correction might be worse compared to closed-loop control. But, any such feed-forward approach to the positioning of an ADC requires accurate knowledge of both the atmospheric dispersion and the instrument optics to reach optimum performance. 

Detailed direct measurements of atmospheric dispersion have been carried out in the infrared by \cite{Skemer2009} and more recently at visible wavelengths by \cite{Wehbe2020}. Both studies used slit-spectroscopy to characterise the dispersion. For this method, the trace map of an observed target is monitored as function of zenith distance. By orienting the direction of dispersion along the slit, the effect of atmospheric dispersion can be resolved from the changing location of the trace on the image plane relative to an observation without dispersion. The major advantage of this approach is that detailed spectral information is obtained in this manner. However, most of the spatial information is removed and thus only a fraction of the telescope field of view can be studied. The removal of spatial information can be circumvented with integral field spectroscopy \citep{Arribas1999}, but no detailed studies characterising dispersion have been done in this manner.

Existing measurements have not yet tested current models of atmospheric dispersion down to the milli arcsecond level necessary to operate ADCs in the optical and near infrared on an ELT. This suggests that an extensive validation of these corrective optics will be necessary during the commissioning phase.

In this work we propose a new approach suitable for the measurement of chromatic PSF elongation in imaging instruments. The proposed method may also be used for the calibration and modification of ADC feed forward control models. Following the concept from \cite{Pathak2016}, we use a spatial filtering technique to introduce broadband artificial speckles to the PSF. Distortion of the speckle shape with increasing differential refraction has been noted in several places \citep{Wang2014}, including in the amateur community\footnote{For example, user KpS recognised the effect in an online forum thread in 2015 at \url{https://www.cloudynights.com/topic/490893-atmospheric-dispersion-corrector-adc-test-images/} [Accessed: 2022-03-23].}, but it has not been derived mathematically. We will do so in section~\ref{section:analysis_speckle_method}. Various spatial filtering techniques are already in use for the projection of artificial speckles next to an observed point source. These speckles act as a photometric and astrometric reference when the central star is obscured in high-contrast imaging systems \citep{Jovanovic2015, Bos2020} or encode information about the observed light, such as wavefront information \citep{Wilby2017}. In these examples either phase modulation by means of a deformable mirror or polarisation modulation by means of a liquid crystal optic was used. In this report, we have opted for amplitude modulation in the form of a specialised Bahtinov mask, primarily because of its mechanical simplicity, design flexibility and low cost. 

We discuss the design of the mask and derive the relevant equations underlying this concept in section~\ref{section:analysis_speckle_method}. Section~\ref{section:observation} provides an overview of the observations we have done. We analyse the obtained data using three different and mostly independent approaches. The speckle based approach, discussed in section~\ref{section:analysis_speckle_method}, uses the diffraction pattern introduced by the mask to geometrically determine the atmospheric dispersion. Section~\ref{section:analysis_simulation_method} describes a second approach that uses forward modelling to mimic the observations. By exploring the dispersion parameter space, we can minimise the difference between observation and simulation. In section~\ref{section:analysis_direct_psf_method}, we illustrate that a direct measurement from the measurement is not a suitable approach to measure atmospheric dispersion. The implications of our results are further discussed in section~\ref{section:discussion} and we finally present our conclusions in section~\ref{section:conclusions}.

\section{Observations}\label{section:observation}
Our observations were done at the Blaauw observatory of the University of Groningen, The Netherlands. The Gratama telescope located at this observatory is a 40 cm Ritchey-Chr\'etien design placed on an equatorial mount, with an SBIG STL6303E CCD detector located at the Cassegrain focus. A summary of several characteristics of the telescope is given in Table~\ref{tab:observatory_information}. 

\begin{figure}
    \centering
    \includegraphics[width=\columnwidth]{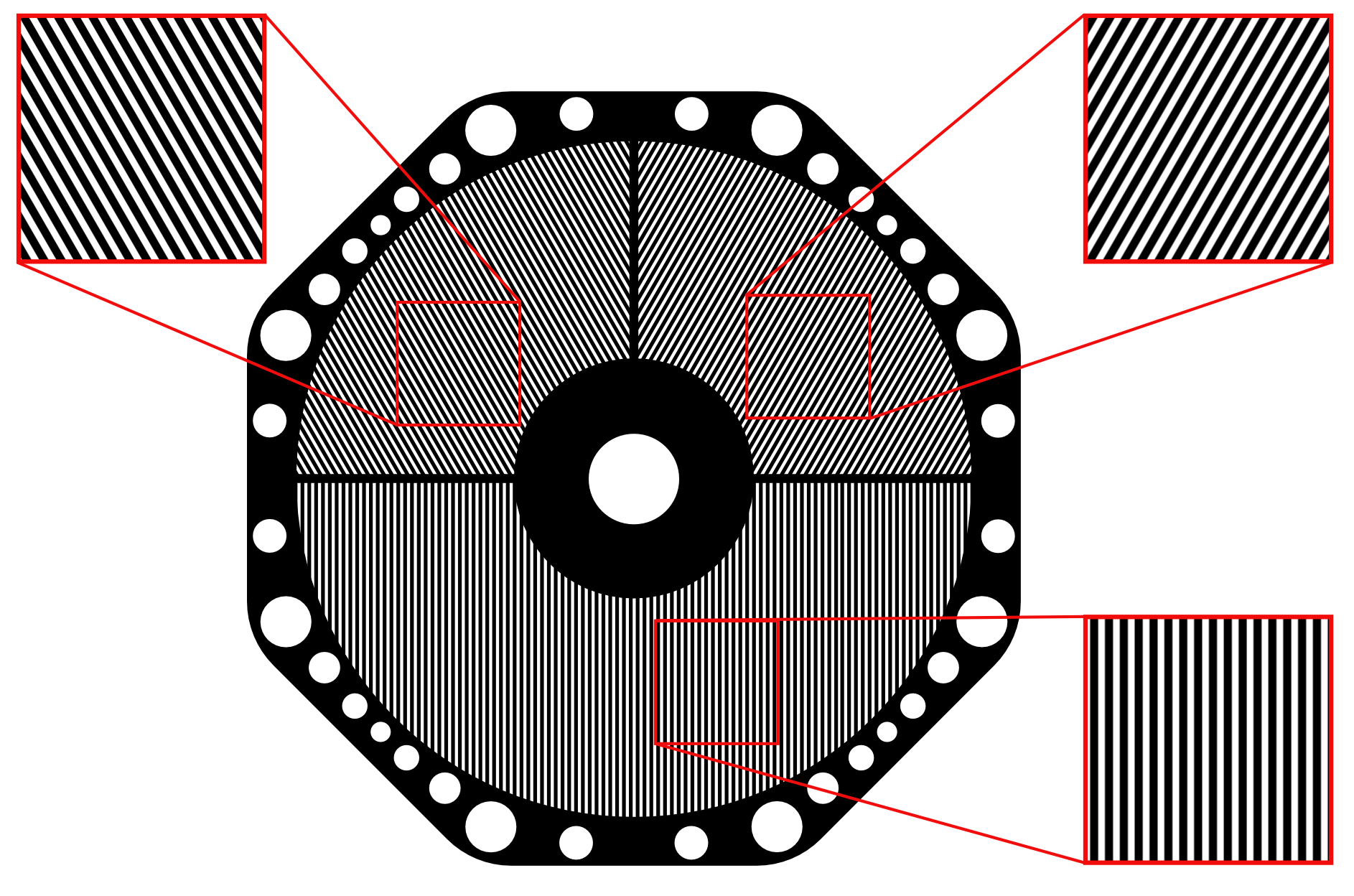}
    \caption{Drawing of the manufactured mask, that was placed on the telescope. Due to the small primary mirror and considerable seeing, 105 lines are necessary to create sufficient distance between the PSF core and the first order diffraction speckles.}
    \label{fig:mask}
\end{figure}

Figure~\ref{fig:mask} shows the mask design. It was laser cut from medium-density fibreboard and painted black. A total of 105 lines over the diameter was chosen, with each line being 2 mm wide, so that the first order diffraction would be well outside the seeing limited disc but not so far as to make source confusion a regular occurrence. The mask was then placed in front of the telescope, attached to the secondary mirror support structure. 

An initial observation was done on the night of 2021 June 13. The average outside temperature was 17$^{\circ}$C and the relative humidity was 79 per cent. The observatory does not support active seeing monitoring, but we estimated the seeing from the full width at half maximum to be around 3~arcsec. Nine stars from the Hipparchos catalogue \citep{Hipparcos1997} were observed at a range of zenith distances, with a particular focus on the lowest altitudes where the atmospheric dispersion is largest. The brightness of the stars was not yet considered as a selection criterion. Also, feed forward autoguiding of the telescope was used during exposures. 

A second observation was done on 2021 October 24. During this night, the average outside temperature was 6$^{\circ}$C with a relative humidity of 85 per cent. This second observation was a done in response to early analyses performed on the data from the first night. We wanted to reduce potential sources of systematic errors and bias in our experiment. We obtained data on eleven stars from the Hipparchos catalogue with similar magnitude ($7.5 < V < 8$) and spectral classification (B9). Also, guide stars were used for the sky tracking of the telescope. 

For each observed target three images were taken in \textit{B} (373 -- 484~nm) and in \textit{V}-band (492 -- 581~nm). All images were reduced using typical flatfield, bias frame and dark current corrections \citep{CCDDataReductionGuide}. 
A detailed overview of the individual observations is given in Appendix~\ref{appendix:observations}.

\begin{table}
 \caption{Summary of the Gratama telescope characteristics}
 \label{tab:observatory_information}
 \begin{tabular*}{\columnwidth}{@{}l@{\hspace*{50pt}}l@{}}
  \hline
  Parameter & Value\\
  \hline
  Telescope design & Ritchey-Chr\'etien\\
  Primary mirror diameter & 400 mm \\
  Focal distance & 3200 mm (f/8) \\
  Obscuration ratio & 0.37\\
  Plate scale & 0.566~arcsec pix$^{-1}$\\
  CCD detector model & SBIG STL6303E \\
  Available filters & \textit{B}, \textit{V}, \textit{R}, \textit{I}, H$\alpha$, H$\beta$, [O\thinspace\small{III}], [S\thinspace\small{II}]\\
  \hline
 \end{tabular*}
\end{table}

\section{Measuring atmospheric dispersion with the speckle method}\label{section:analysis_speckle_method}
\subsection{Method description}\label{subsubsection:speckle_method_method}

Bahtinov masks are a simple type of diffraction mask and are used regularly on smaller telescopes to aid in focus verification. An optimal focus is reached when the diffraction spikes introduced by the mask intersect on top of an observed star, see \cite{Zandvliet2017} for a case study. 

The primary design parameters of the diffraction mask are the line density and the line orientation. The first determines the distance between the first order diffraction speckles and the central star. The second determines the direction in which the diffraction speckles appear. By combining multiple line orientations in the mask, multiple sets of diffraction speckles are created at angular positions corresponding to the line directions. For this work, we have chosen to adhere to the classical Bahtinov design with three separate pupil zones, because of its design simplicity and ease of production. Two of the six diffraction speckles created by the mask would be brighter, as these corresponded to the larger pupil zone on the mask. More optimal mask designs may exist, such as a superposition of lines with different orientations or a pattern of repeating sub-apertures. These are not yet investigated in detail.

Now, we argue that a mask of this type, designed with sufficient lines over the mask diameter, can be used to measure the presence of chromatic elongation of the PSF.

Because the diffraction mask resembles an amplitude grating, we begin the derivation from the grating equation
\begin{equation}
    d \left(\sin{\theta_i} - \sin{\theta_m}\right) = m\lambda, \label{eq:grating_equation}
\end{equation}
where $d$ is the grating period, $\lambda$ is the wavelength of the light, $\theta_i$ and $\theta_m$ are the angles of the incoming and outgoing ray relative to the grating normal. Finally, $m$ is the integer denoting the order of diffraction.

For a mask placed at the entrance pupil of a telescope, we may apply the small angle approximation, $\sin\theta_m \approx \theta_m$, and say that $\theta_i \approx 0$. Furthermore, we can rewrite the grating period $d$ in terms of the entrance pupil diameter $D$ and the number of lines $N$ over this diameter. Then,
\begin{equation}
    \theta_m  \approx N \frac{m\lambda}{D}. \label{eq:speckle_distance}
\end{equation}

\begin{figure}
    \centering
    \includegraphics[width=0.7\linewidth]{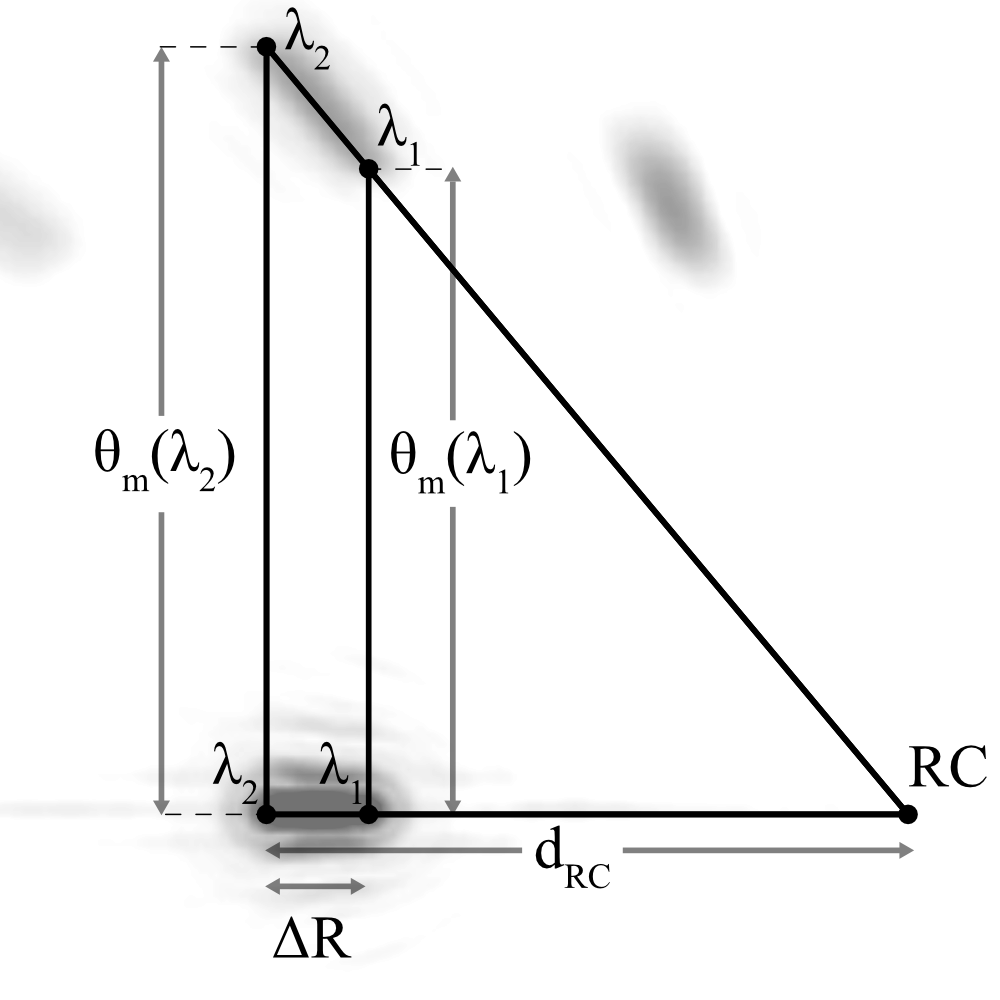}
    \caption{The geometry of a point source experiencing some amount of dispersion $\Delta R$, as it is observed with a diffraction mask. The angular distance between the PSF and the diffraction speckle is given by $\theta_m(\lambda)$. Due to the dispersion, the polychromatic diffraction speckle pattern will point towards the radiation centre ($\mathrm{RC}$). The distance from the PSF centre at $\lambda_2$ to the radiation centre is defined as $d_{\textnormal{\tiny RC}}$.}
    \label{fig:speckle_method}
\end{figure}

When a point source experiences differential refraction at wavelengths $\lambda_1$ and $\lambda_2$, we may evaluate equation~\eqref{eq:speckle_distance} at each wavelength independently. From the resulting geometry, illustrated in Fig.~\ref{fig:speckle_method}, we find that
\begin{equation}
    d_{\textnormal{\tiny RC}} = \left( \frac{\lambda_2}{\lambda_2 - \lambda_1}\right)\Delta R. \label{eq:speckle_equation}
\end{equation}

The angular position of the speckles relative to the zeroth diffraction order, the PSF core, depends on the orientation of the grating. If multiple grating directions are included in the mask design, three in case of a classical Bahtinov mask, then speckles appear at six angular positions relative to the central PSF. Indeed, all diffraction speckles will point towards the same intersection point if the telescope is in focus. Following \cite{Pathak2016}, we will call this intersection point the radiation centre. The distance from the radiation centre to the location where a ray with wavelength $\lambda_2$ would hit the focal plane is given by $d_{\textnormal{\tiny RC}}$. This distance is a direct magnification of the differential refraction, $\Delta R$, experienced by the observed target.

\begin{figure}
    \centering
    \includegraphics[width=\linewidth]{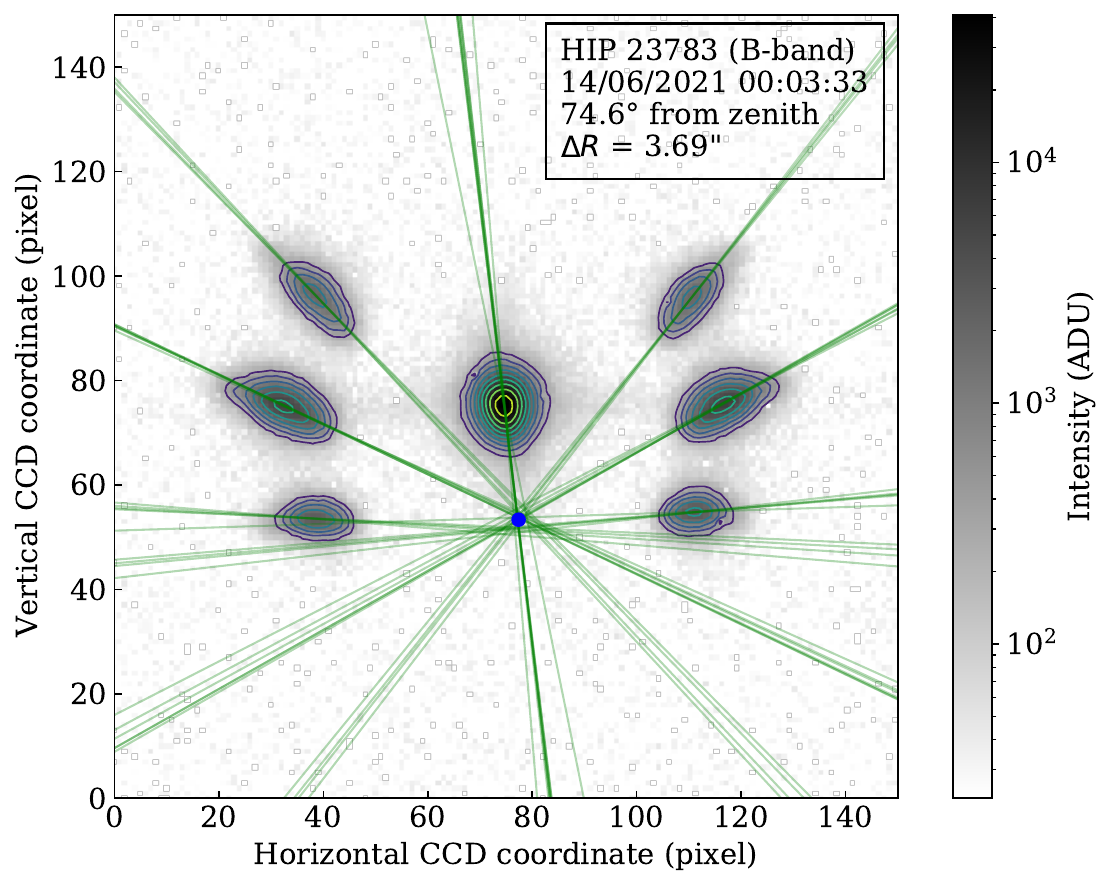}
    \caption{The radiation centre (blue dot) is found from a least squares intersection of the lines through the major axes of the best-fitting ellipses of the contour plot.}
    \label{fig:applied_speckle_method}
\end{figure}

To infer the atmospheric dispersion from the speckles and the PSF core, we must find the intersection point to which all speckles point. For this we perform the following steps on the image. First, we create a contour plot of the image at ten intensity levels with a signal to noise larger than two, equally spaced in log-space. We select only the contours that are closed and have more than eight nodes. This ensures that only the speckles and the central star are included and most noise sources are removed. Next, we perform a non-iterative least squares fit of an ellipse to each contour \citep{halir1998} and use the information of these ellipses to draw lines through the major axes of all of them. Then a least squares intersection of the lines is determined to find the radiation centre \citep{Traa2013}. This process is explained in Appendix~\ref{appendix:least_squares_intersection}. Figure~\ref{fig:applied_speckle_method} illustrates this process as applied to our \textit{B}-band image of HIP~23783.

As our next step, we fit a rotated elliptical Moffat distribution to the PSF core and determine its centroid. The intensity profile of a non-rotated Moffat profile is described by

\begin{align}
    M(x,y) = I_{\textnormal{\tiny bg}} + I_0 \left( 1 + \frac{x^2}{\alpha_x^2} + \frac{y^2}{\alpha_y^2}\right)^{-\beta},\label{eq:moffat_psf}
\end{align}
where $I_{\textnormal{\tiny bg}}$ is the background intensity, $I_0$ the signal amplitude and $\beta$ the shape parameter \citep{Moffat1969}. The scaling parameters $\alpha_x$ and $\alpha_y$ contain information about the full-width at half-maximum (FWHM),
\begin{align}
    \alpha_i = \frac{\mathrm{FWHM_i}}{2\sqrt{2^{1/\beta}-1}},
\end{align}

Now, the profile is positioned at any angle through a rotation and translation of the coordinate frame,
\begin{align}
    \begin{pmatrix}x \\ y\end{pmatrix} = 
    \begin{pmatrix}
    \cos\theta & \sin\theta \\
    -\sin\theta & \cos\theta
    \end{pmatrix} \begin{pmatrix}
    x' - x_0 \\
    y' - y_0
    \end{pmatrix}.\label{eq:moffat_rotated}
\end{align}
This coordinate transformation is only used to find the best location and orientation of the PSF core and is therefore independent of the locations of the speckles.

We have to know the location of the PSF centroid for a wavelength $\lambda_2$ to apply equation~\eqref{eq:speckle_equation}. This is not straight forward for seeing limited observations. Therefore, we modify the expression slightly. We can assume that the centre of the Moffat profile, equation~\eqref{eq:moffat_psf}, will be very close to where the central wavelength would hit the focal plane. If we then assume that the atmospheric dispersion behaves linearly over the passband, we can write the central wavelength as $\lambda_c = (\lambda_1 + \lambda_2)/2$ and measure $d_{\textnormal{\tiny RC}} = d_c + \Delta R/2$, with $d_c$ being the distance from the radiation centre to the Moffat PSF centroid. This then leads to the modified expression
\begin{equation}
    \Delta R = 2 \left(\frac{\lambda_2 - \lambda_1}{\lambda_2 + \lambda_1}\right) d_c. \label{eq:speckle_equation_modified}
\end{equation}

\begin{figure}
    \centering
    \includegraphics[width=\columnwidth]{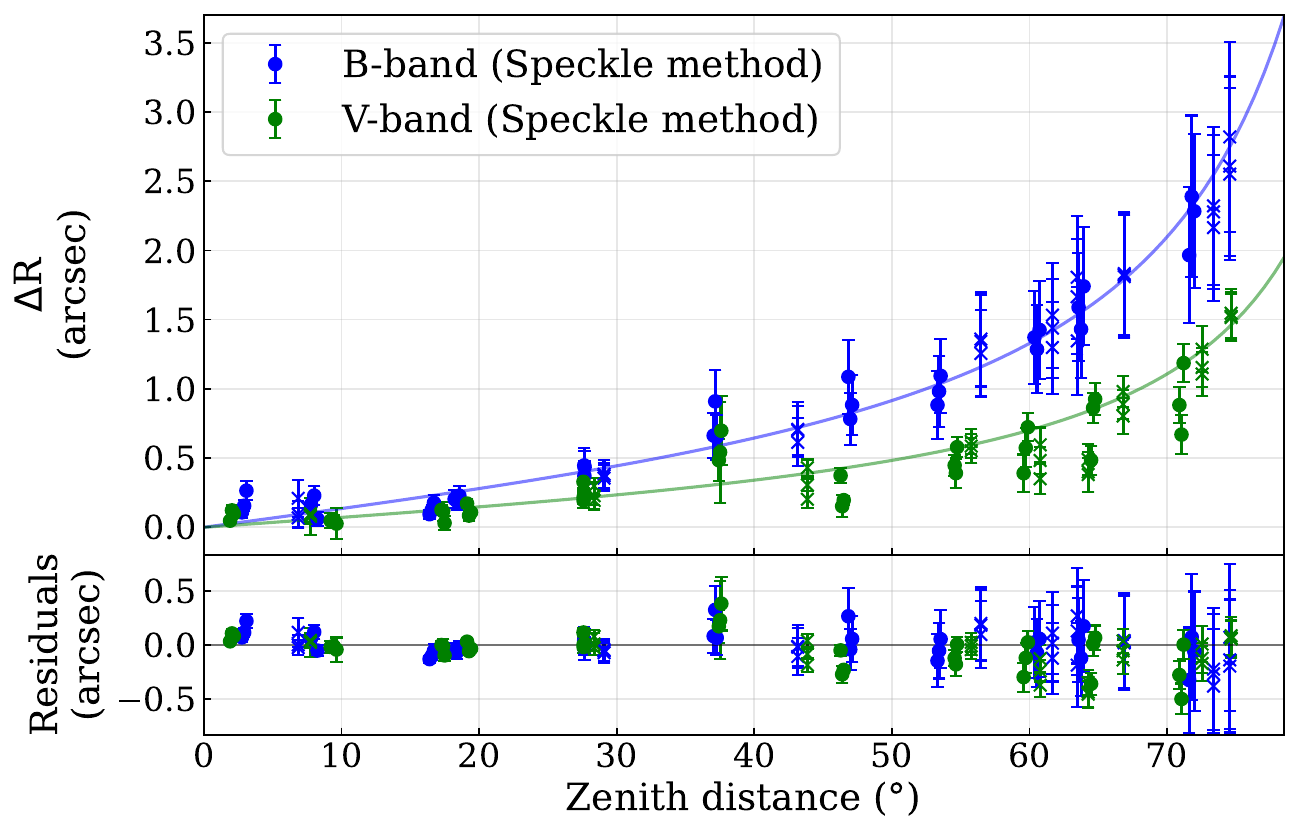}
    \caption{The measured atmospheric dispersion in arcseconds using the artificial speckles introduced by the diffraction mask. Data points from the first observing run are denoted by a cross, while data point from the second observation are given by a filled dot. We see good agreement with the expected atmospheric dispersion of equation~\eqref{eq:atmospheric_dispersion} (solid lines). The bottom panel shows the difference between the measured and expected value.}
    \label{fig:speckle_results}
\end{figure}

Now, we can apply these steps to all our observations. The measured atmospheric dispersion as a function of zenith distance is presented in Fig.~\ref{fig:speckle_results}. We find that our measurements clearly reveal the expected trend and are in good quantitative agreement with equation~\eqref{eq:atmospheric_dispersion} to within 0.5~arcsec. Furthermore, we do not observe a significant difference in the residuals between the two different observing nights.

\subsection{Uncertainties}\label{subsection:specklemethod_uncertainties}
The uncertainties presented in Fig.~\ref{fig:speckle_results} generally increase with the magnitude of the atmospheric dispersion. These consist of static contributions from the telescope spectral response and the plate scale, as well as variable contributions from the location of the PSF core centroid and the location of the radiation centre. In this subsection we provide an overview of the assumed or determined error contributions.

All contributing parameters were derived from equation~\eqref{eq:speckle_equation_modified}, including a conversion of $d_c$ to pixel coordinates,
\begin{equation}
    d_c = p \sqrt{(x_{\textnormal{\tiny RC}} - x_c)^2 + (y_{\textnormal{\tiny RC}} - y_c)^2}. \label{eq:dc_explicit}
\end{equation}
Here $(x_c, y_c)$ is the location of the PSF core centroid and $(x_{\textnormal{\tiny RC}}, y_{\textnormal{\tiny RC}})$ are the coordinates of the radiation centre. The plate scale in arcseconds per pixel is denoted by $p$.

\begin{figure}
    \centering
    \includegraphics[width=\linewidth]{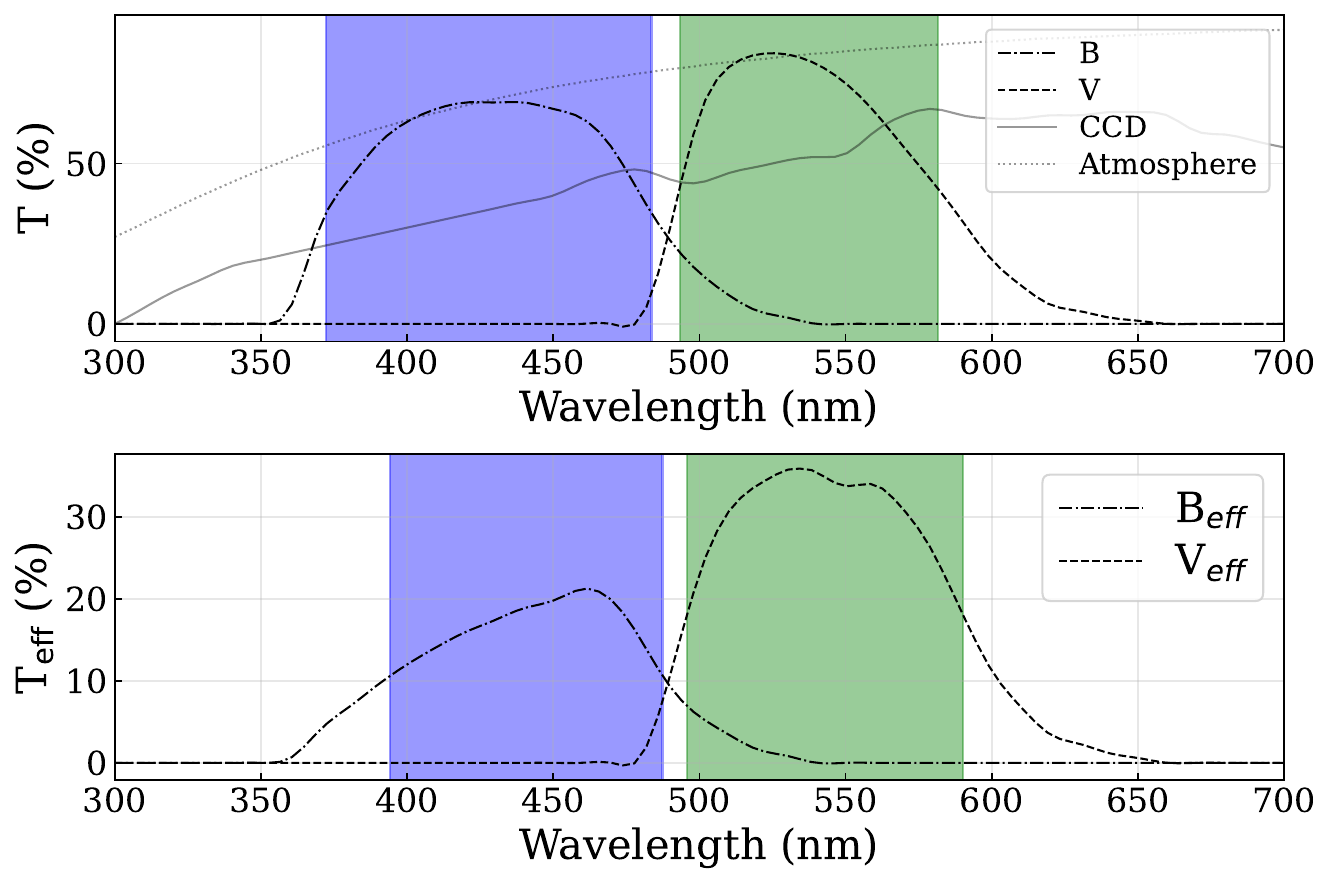}
    \caption{The half-maximum transmission level of the separate filters differs if the quantum efficiency of the CCD and the atmospheric transmission is included. This creates a non-negligible uncertainty in the wavelengths for which the atmospheric dispersion is measured.}
    \label{fig:effective_transmission}
\end{figure}

Because atmospheric dispersion is strictly defined as the differential refraction between two different wavelengths, we must choose these wavelengths based on the spectral content of the observation. Ideally, the combined response of the telescope, filters and atmosphere form a top-hat distribution where the differential refraction between the cut-on wavelength and the cut-off wavelength would directly determine the elongation due to dispersion. This is not a realistic assumption in most cases. Therefore, we have used the half-maximum transmission points of the effective telescope transmission curve as the cut-on and cut-off wavelength. Note, however, that the half-maximum transmission points are different depending on the number of effects included, as illustrated in Fig.~\ref{fig:effective_transmission}. If we include the filter transmission curves, the normalised atmospheric transmission and the quantum efficiency of the CCD, then we find that the cut-on and cut-off wavelengths of the filters must be shifted by $+22$~nm and $+4$~nm for \textit{B}-band and $+3$~nm and $+9$~nm for \textit{V}-band, relative to the original filter definitions. The results presented in Fig.~\ref{fig:speckle_results} thus show the elongation of the PSF due to atmospheric dispersion in an effective passband between 395 and 488~nm in \textit{B} and between 495 and 590~nm in \textit{V}.

Because the transmission curves included in this estimate were not experimentally verified, we include these numbers as a systematic uncertainty in the measurement. The stellar classification of the observed target and the reflectivity curves of the telescope mirrors were considered lower order contributions, and therefore not included in this analysis. This is also reflected in the negligible difference between the measured dispersion of the two nights.

The second systematic error term is the plate scale. The plate scale of the Gratama telescope was obtained from archived observations of standard field SA32SF4 \citep{Landolt2013}. We supplied a list of the image coordinates of the thirty brightest objects in the field to the web version of Astrometry.net \citep{Lang2010}. Both \textit{B} and \textit{V} images resulted in the same plate scale value of 0.566~arcsec pix$^{-1}$. A characterisation of the telescope done in 2009 found a standard deviation around this value of 0.001~arcsec \citep{Janssen2009}. We have adopted this value for this work.

Next, the uncertainty in the radiation centre location is determined empirically for each image. We define the $1\sigma$ uncertainty to be equivalent to the radius of a circle in which 68 per cent of the lines through the speckles, multiplied by their confidence level, intersects or hits this circle. We find that this term increases significantly as $d_c$ increases and can be as large as 0.7 arcseconds at the largest zenith distances.

The final parameter uncertainty is found from the location of the PSF core. This term is determined from the covariance matrix of the least squares fit of the elliptical Moffat profile, described by equations~\eqref{eq:moffat_psf}-\eqref{eq:moffat_rotated}, and is generally small.

We apply classical error propagation on equation~\eqref{eq:speckle_equation_modified} to find the final combined uncertainty in $\Delta R$,
\begin{equation}
    \sigma_{\Delta R}^2 = \sum^N_{i=0} \left( \frac{\partial (\Delta R)}{\partial i} \sigma_i \right)^2. \label{eq:error_prop}
\end{equation}

Overall, the uncertainties range from about 0.05 arcseconds near zenith to at most 0.7 arcseconds in \textit{B}-band at the largest observed zenith distances.

\section{Forward model based approach}\label{section:analysis_simulation_method}
The Gratama telescope is a simple two mirror telescope with the detector placed directly in the Cassegrain focus. Therefore, it is relatively easy to model the observations using Fourier optics. This aids in our analysis and provides an independent check of our physical understanding of the observed images.

The diffraction limited image response of the telescope is given by the Fraunhofer diffraction integral of the complex field distribution at the telescope entrance pupil \citep{Goodman2005}. The intensity distribution at the telescope focal plane can be written as
\begin{equation}
    I(x,y,\lambda) = | \mathcal{F} \left\{ E_{\textnormal{\tiny in}}(\xi, \eta, \lambda) \right\}|^2, \label{eq:FraunhoferApprox}
\end{equation}
where the Fourier transform operation, here denoted by $\mathcal{F}$, must be evaluated at frequencies $x/(\lambda z)$ and $y/(\lambda z)$. The focal length of the telescope is in this case equivalent to the propagation distance $z$.
The Bahtinov mask we used defines the entrance aperture of the telescope and forms the amplitude of the complex input field $E_{\textnormal{in}}$, while the atmospheric refraction at $\lambda_i$ is represented as a wavelength dependent tilt of the complex phase.

To model the long exposure PSF, we simulate the atmospheric seeing by multiplication of the optical transfer function of the telescope and that of the atmosphere, assuming a Von-Kármán power spectrum for the atmospheric turbulence. See \citet{Fetick2019} for a more explicit treatment.

The polychromatic image is treated as the sum of the images at $N$ separate wavelengths, weighted by the effective transmission at that wavelength and scaled to the desired amplitude $I_0$.
\begin{equation}
    I(x,y) = I_0 \sum^{N}_{i=1} w_i I(x,y,\lambda_i)
\end{equation}

Then, the output intensity distribution is binned into the spatial resolution elements of the CCD detector to create the final image.

To compare the spatial overlap between the simulated and observed images, we use the peak value of a normalised two-dimensional cross-correlation. We denote this value by $\eta$. Considering the intensity only, it is defined by

\begin{align}
    \eta &= \displaystyle\int I_{\textnormal{obs}}I_{\textnormal{sim}}dA &&\textnormal{ where }  0 \leq \eta \leq 1.\label{eq:efficiency}
\end{align}

\begin{figure*}
    \centering
    \includegraphics[width=\linewidth]{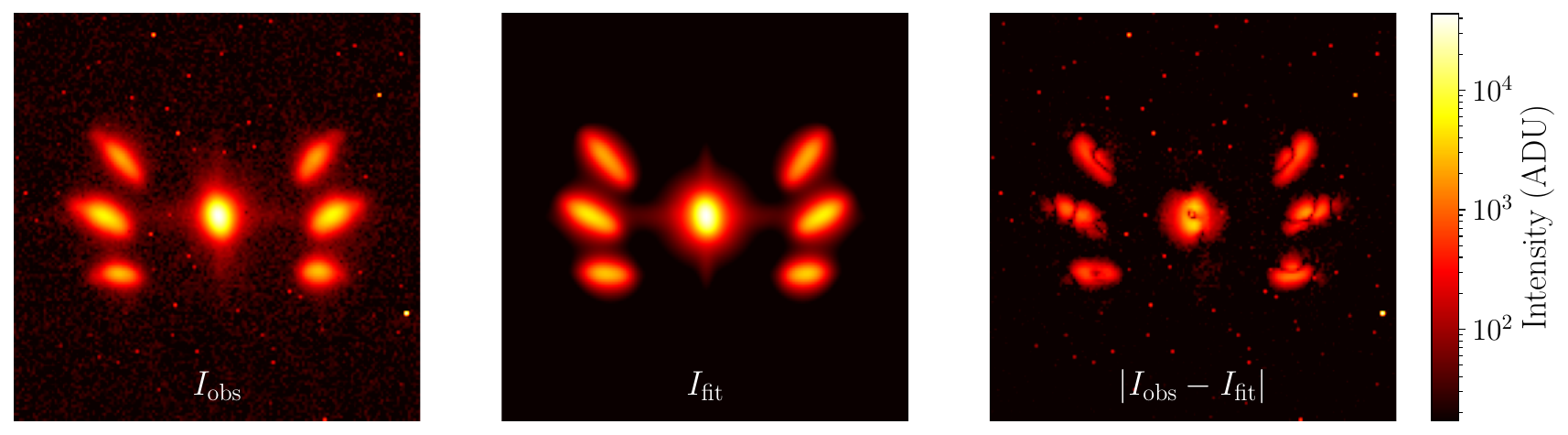}
    \caption{Comparison of our \textit{B}-band observation of HIP 23783 (left), the simulated image (middle) and the difference image (right). All three images use the same logarithmic intensity scale. Here, $\eta=0.942$.}
    \label{fig:simulation_comp}
\end{figure*}

\begin{figure}
    \centering
    \includegraphics[width=\columnwidth]{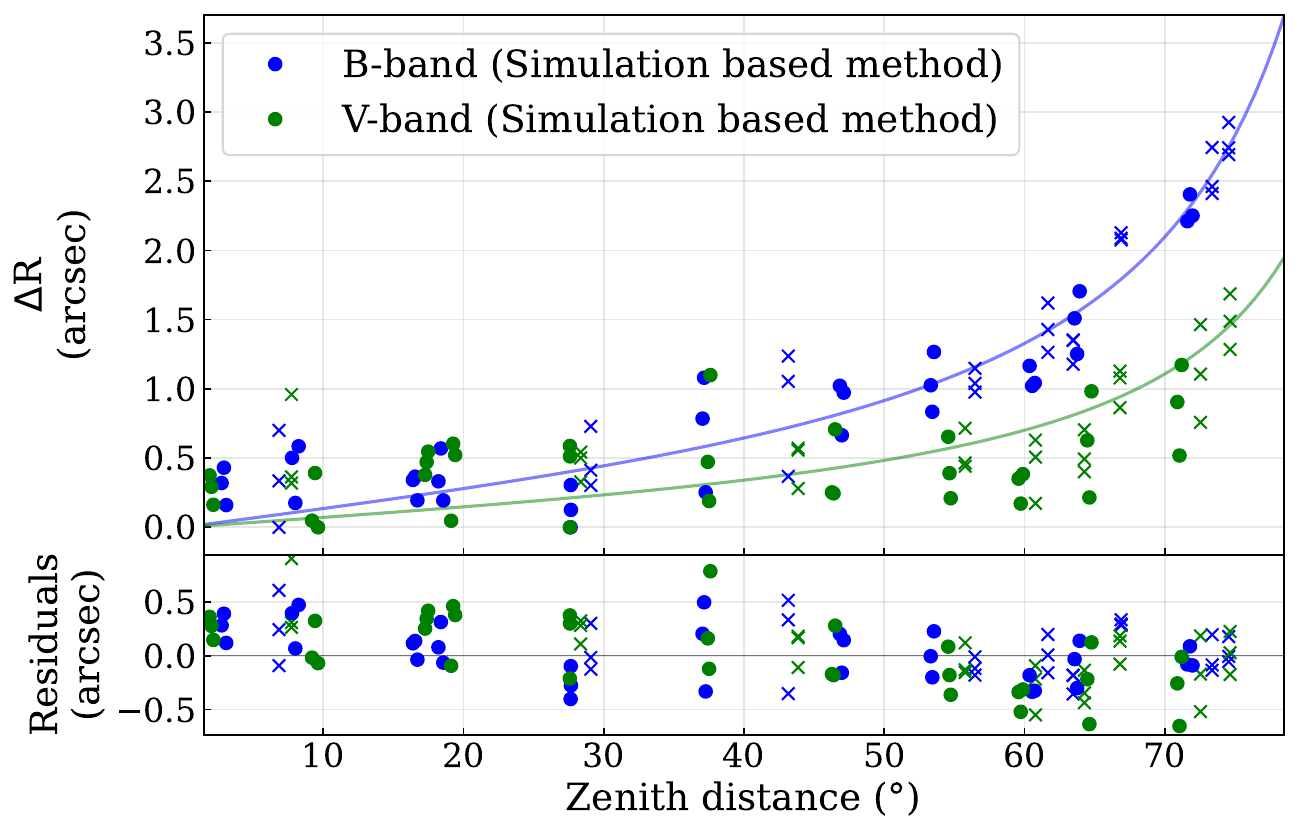}
    \caption{We have estimated the atmospheric dispersion by maximising the match between simulated images and the real images. Data points from the first night of observations are denoted by a cross, while data point from the second night are given by a filled dot. The bottom panel shows the residuals between the estimated dispersion and the expected dispersion at the respective zenith angle.}
    \label{fig:simulation_results}
\end{figure}

We apply our knowledge of the telescope, the mask, the filter and CCD transmission curves and CCD properties to an image calculated using the image formation theory described above. The maximum intensity and the background intensity are extracted from the data to obtain a closer match with the real images. As the observatory does not have a dedicated seeing monitor, we use equations~\eqref{eq:moffat_psf}-\eqref{eq:moffat_rotated} to estimate the seeing in the image from the FWHM of the PSF core. 

The computations are done on a 3001$\times$3001 grid and repeated for 15 wavelength slices over the passband. The resulting images at the separate wavelengths are summed and scaled to the desired intensity. The resulting image is then binned into the 9 \textmu m size of the STL6303E CCD pixels. The dispersion vector, containing the magnitude and the direction of the dispersion, is the free parameter for which a maximisation routine is used with equation~\eqref{eq:efficiency} being the function of merit. 

One comparison of the original observation and the best fit simulated image is shown in Fig.~\ref{fig:simulation_comp}. Application of the above to all our observations give the results shown in Fig.~\ref{fig:simulation_results}.

\begin{figure}
    \centering
    \includegraphics[width=\columnwidth]{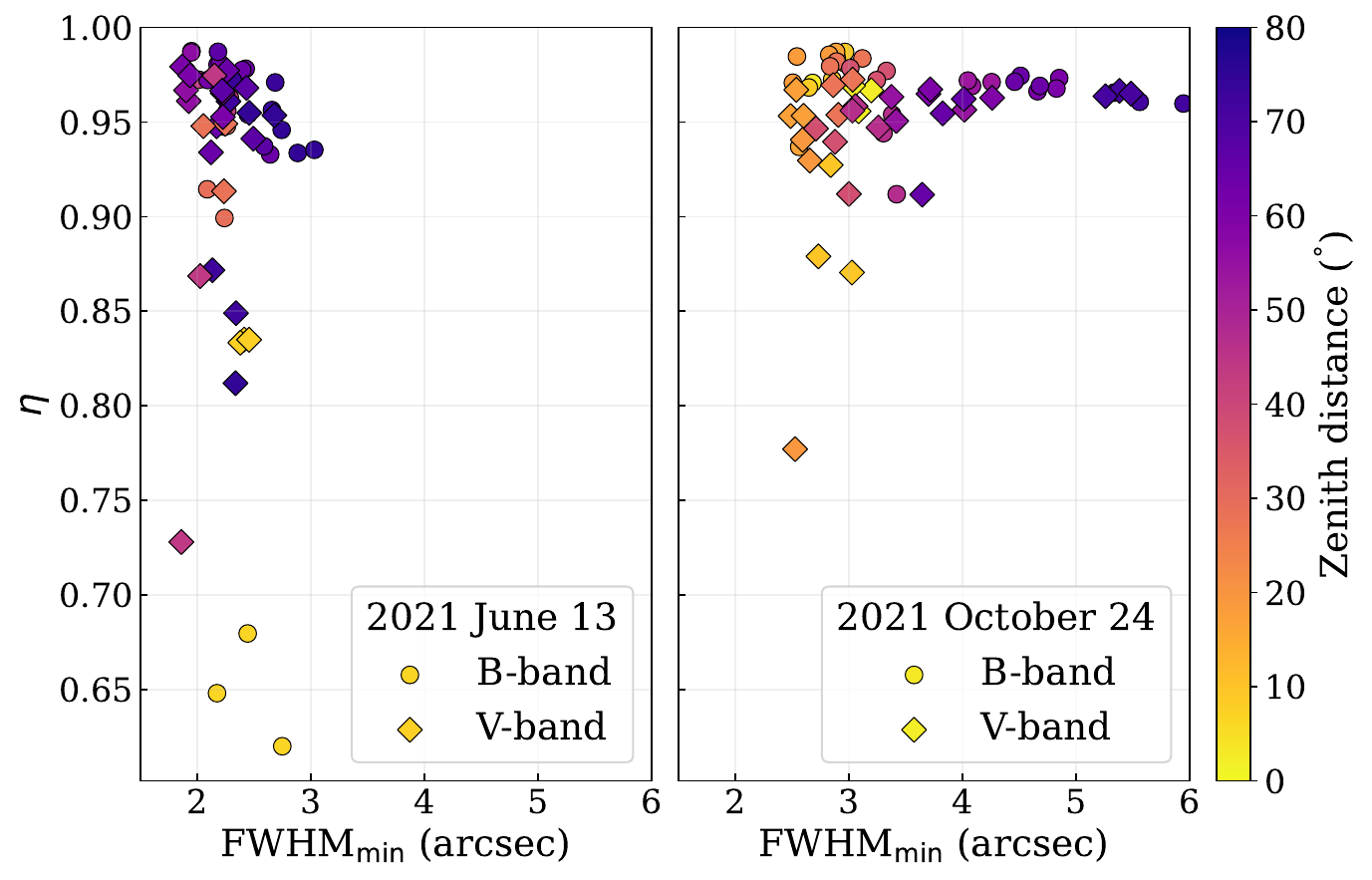}
    \caption{A comparison of the cross-correlation term $\eta$ and the observed seeing, the FWHM of the PSF in the direction perpendicular to elongation, suggests that $\eta$ decreases when both the zenith distance and observed seeing are small. To illustrate the different observing conditions for the first and second observing night, we have split the figure. Note that the y-axis starts at $\eta = 0.6$.}
    \label{fig:coupling_vs_zenith}
\end{figure}

We find that this method is usually able to retrieve the atmospheric dispersion to within 0.5~arcsec. At larger zenith distances, we can clearly discern the two passbands. We do, however, see more scatter of the data points compared to the speckle method. Also, at zenith distances lower than 30$^{\circ}$, this method fails to find the correct magnitudes. Instead, it tends to overestimate the PSF elongation, where little is expected. Inspection of the cross-correlation term $\eta$, the observed seeing and the zenith distance, reveals that the fit quality deteriorates when both the observed seeing and zenith distance are small, as shown in Fig.~\ref{fig:coupling_vs_zenith}. If we assume that this effect is neither a function of seeing nor zenith distance, then the decrease in $\eta$ can be explained as a relative dominance of this non-modelled error term. Telescope tracking errors, vibrations and other effects that are excluded from the model, may be the culprit. This hypothesis is strengthened by a visual inspection of the three \textit{B}-band images taken of HIP~73005, at 6.9$^{\circ}$ from zenith. These show clear variations in the shape of the PSF. The integration times were sufficiently long that this variation can not easily be explained by atmospheric turbulence. Thus a movement of the telescope, relative to the star, is more likely.

\section{Direct measurement of the PSF core}\label{section:analysis_direct_psf_method}
\subsection{Method description}\label{subsection:direct_psf_method}
To illustrate the advantages of the methods proposed in the previous sections over a direct analysis of the PSF core, we succinctly discuss such an analysis on our data.

To fit the elongation of PSF core, we use a Gaussian distribution that gets dispersed symmetrically by a length $i$ along the axis of elongation,
\begin{equation}
    \begin{aligned}
        G_s (x,i) &= \int^{\frac{i}{2}}_{-\frac{i}{2}} \exp{\left[-\frac{1}{2}\left(\frac{x - a}{\sigma}\right)\right]} da \\
        &= A_x \left[ \mathrm{erf}\left(\frac{i/2 - x}{\sigma \sqrt{2}}\right) + \mathrm{erf}\left(\frac{i/2 + x}{\sigma \sqrt{2}}\right) \right]. 
    \end{aligned}\label{eq:shifted_gauss}
\end{equation}
Here $\mathrm{erf}(x)$ denotes the error function, $\sigma$ the standard deviation of the original Gaussian distribution and $A_x$ is an undefined scaling factor to scale the dispersed Gaussian to the desired amplitude.

An analytical derivation of the two-dimensional version of $G_s$ was not done. However, we determined that equation~\eqref{eq:shifted_gauss} can be evaluated in both $x$ and $y$, separately, by assuming a square separable profile function. Then, a multiplication of the two, followed by another rescaling of the amplitude factor, gives the expected two-dimensional shape.

\begin{figure}
    \centering
    \includegraphics[width=0.85\columnwidth]{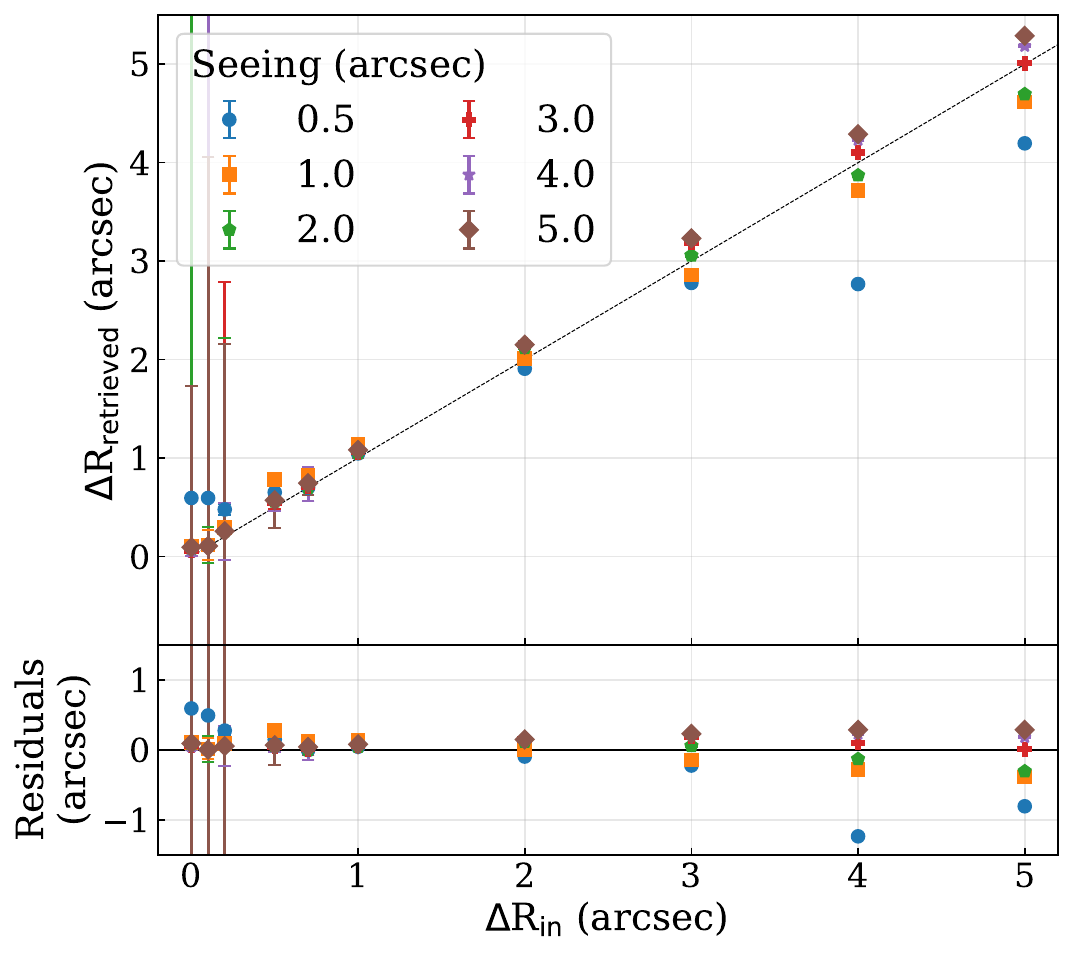}
    \caption{Simulated images were used to verify the retrieval of the atmospheric dispersion from the PSF core. When the atmospheric dispersion becomes less than the pixel scale, the fit is no longer reliable. For dispersion levels between one and four arcseconds the fit yields the correct value to within a few tenths of an arcsecond. There is a small dependence on the atmospheric seeing.}
    \label{fig:verification_direct_psf_method}
\end{figure}

Before we turned to the acquired images, we verified the method with the simulation framework described in the previous section. Figure~\ref{fig:verification_direct_psf_method} illustrates that we can recover a given PSF elongation to within a few tenths of an arcsecond. We observe some variation between the given and recovered elongation when the PSF is nearly undersampled, here when the seeing is 0.5~arcsec. At worse seeing levels the recovery of the dispersion works as expected, although a slight divergence can be observed as the elongation grows larger.

After this numerical verification of the method, we now perform the analysis on the real data.

\begin{figure}
    \centering
    \includegraphics[width=\columnwidth]{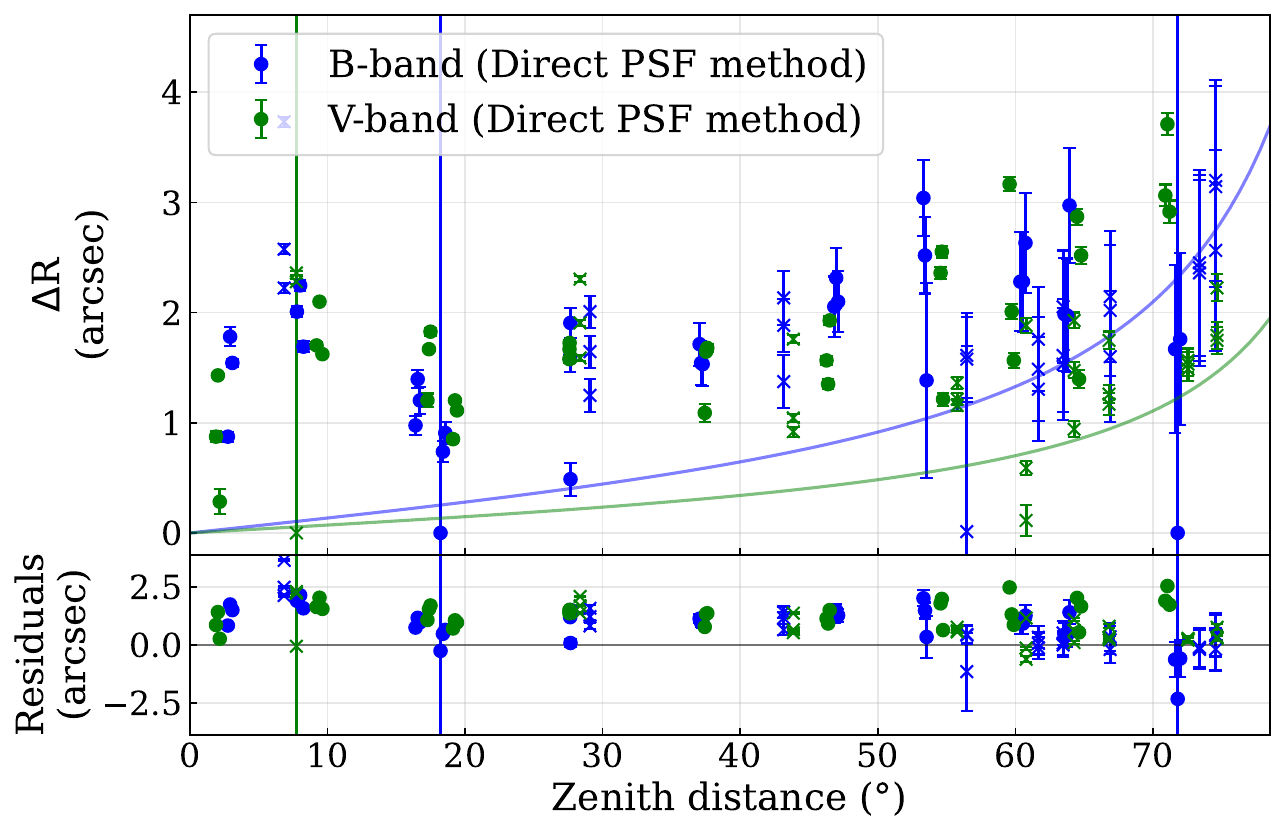}
    \caption{The atmospheric dispersion as measured directly from the star. \mbox{Using} the dispersed Gaussian function we find the PSF elongation as a function of zenith distance. Again, crosses mark data points from the first observing run and filled dots denote data from the later observation. The bottom panel shows the residuals between the measured and expected atmospheric dispersion.}
    \label{fig:direct_psf_results}
\end{figure}

As expected, the results shown in Figure~\ref{fig:direct_psf_results} are not as convincing as in our other analyses. Most notable is the high systematic offset of roughly 1.5~arcsec in the measured elongation, that dominates the elongation up to zenith distances of approximately 45$^{\circ}$. Even after this point it is difficult to discern any clear trends, except for an increase in the variation of the measured elongation. This again hints at a temporal instability of the telescope or atmosphere that results in a smearing of the PSF, as previously discussed in section~\ref{section:analysis_simulation_method}.

\subsection{Uncertainties}\label{subsection:direct_psf_uncertainties}
We estimated the uncertainty, shown by the error bars in Figure~\ref{fig:direct_psf_results}, as follows. The covariance matrix of the least squares minimisation routine determines the uncertainty in the Gaussian shift $i$. Together with the plate scale uncertainty of 0.001~arcsec pix$^{-1}$, this determines the error in the dispersion detected on the CCD. However, the dominant source of error is originating from the wavelength dependence. As described in section~\ref{subsection:specklemethod_uncertainties}, the passband limits change when taking the full spectral response into account. Next, the FWHM of the elongated PSF is a function of the FWHM of the PSF at the observed wavelength, the diffraction limit and the atmospheric dispersion \citep{Martinez2010, vandenBorn2020}. Finally, it should be noted that we did not include an additional error term for the telescope pointing accuracy. From our results, we would expect this to be on the order of 1-2 arcseconds for this particular telescope.
Assuming independence, we estimate the total uncertainty as a root sum square of the errors as derived from their analytic expressions.

\section{Discussion}\label{section:discussion}

We discern no clear difference in the obtained dispersion between the observations of 2021 June 13, where no guide stars were used for the telescope tracking, and 2021 October 24, where guide stars were used. Our hypothesis is that the large measured elongation at low to moderate zenith angles is a result of a limited tracking accuracy of the telescope mount.  To investigate this hypothesis a bit further, we have looked at several VIMOS observations \citep{VIMOS2003} from the ESO Scientific Archive. VIMOS was chosen as it did not incorporate an ADC during operations, while many more recent instruments do, and it should offer better telescope stability compared to the Gratama telescope. In each of the images, we selected approximately one hundred appropriate point sources and measured the PSF elongation using the direct PSF method described in the previous section. Consistently, the mean and median measured elongation in the images exceeded the value that would be expected from atmospheric dispersion alone.

In principle, the fixed contribution to the elongation that result from optical aberrations can be removed with sufficient data or a good understanding of the instrument design. But other causes, such as guiding errors, wind shake or instrumental vibrations may be more difficult to disentangle from imaging data only \citep{spyromilio2021}. At large zenith distances the atmospheric dispersion becomes the dominant effect and there the relative measurement accuracy improves. When the ELTs will come online, it may be interesting to reconsider this method. However, when accuracies below the diffraction limit are required, the speckle method provides a more robust approach.

There are several advantages of the speckle method. First of all, it has shown to be less sensitive to systematic errors of a non-chromatic origin, compared to a direct measurement of the PSF core. Assuming that the diffraction mask can be placed at the optical pupil of an instrument, and that it can be used concurrently with a spectral bandpass filter, then the speckle method can be used to verify the performance of the atmospheric dispersion corrector before and after the scientific observation. For the observation of interest, the mask would then be removed from the light path. The performance characterisation can be done for any position in the field, because of the masks location in the pupil. Note, however, that the impact of field dependent effects, such as wavefront aberrations, geometric distortions and anisoplanatism, were not investigated thoroughly for this report. One example we can mention is that the edges of the image plane may not be in focus in a system with considerable field curvature. Then the lines drawn from the artificial speckles will not intersect in a single point. Instead, they intersect at multiple distinct points along the direction of dispersion. The distance between these intersection points is a function of both the dispersion and the defocus. While this specific problem can be resolved by using a rotation symmetric mask design, other issues might arise for other types of aberrations.

The forward modelling approach can be a decent method for the estimation of dispersion present in the system. The post-observation reconstruction of the image is not easy, but significant effort is being put into developing the necessary tools for current and future instruments (for example \cite{Gilles2018, Jolissaint2018, Massari2020, Simioni2020}). The addition of (residual) atmospheric dispersion as a parameter in these tools could offer a good way to measure or monitor the dispersion at large observatories. However, it requires significant effort and computation time before results can be obtained. Furthermore, our results indicate that this method can overestimate the dispersion when little is present, if one or more lower order effects are not included in the model. These effects then only become apparent when the dispersion and atmospheric seeing are small.

While no major deviations from standard descriptions of atmospheric dispersion were expected for this work, we do not yet know if existing models are sufficiently accurate to satisfy the requirements of the upcoming extremely large telescopes. The speckle method developed in section~\ref{section:analysis_speckle_method} offers opportunities to test the atmospheric dispersion models on larger telescopes and enables the development of high precision models for feed forward correction, when no closed loop control is possible. For seeing limited observations, the presented results can be improved upon by using a larger telescope located at a better observing site. The presence of adaptive optics facilities, frequent plate scale calibrations and much more stringent requirements on the end-to-end effective transmission at many major observatories will further improve the accuracy of the measurements.

\section{Conclusions}\label{section:conclusions}
In this work we have presented and compared three different methods to measure atmospheric dispersion from imaging data. We have shown that the addition of a diffraction mask spatially filters the image, thereby magnifying the dispersion by a factor $\lambda_2/(\lambda_2 - \lambda_1)$. This makes it easier to detect its presence. We tested the concept on the 40 cm Gratama telescope and found that our results agreed with the expected dispersion to within 0.5~arcsec, even with >2~arcsec atmospheric seeing.

A forward modelling approach, that incorporated our knowledge about the telescope and the observing conditions, provided an independent check for the geometric approach of the speckle method. We fitted the atmospheric dispersion magnitude and direction as the only free parameters of the simulated images. The results agreed with the expected levels of atmospheric dispersion to within 0.5~arcsec, but had larger variation in the obtained magnitudes compared to the speckle method. At small zenith distances, where the dispersion should be minimal, this method overestimated the dispersion. This can be explained by temporal instabilities of a non-chromatic origin, such as telescope vibrations or sky tracking issues.

A third method was discussed to illustrate that a characterisation of atmospheric dispersion from imaging data is not feasible without a diffraction mask. Using a description of a dispersed Gaussian distribution, the PSF elongation was determined from the star image directly. Numerical simulations suggested that accuracies of $\leq 1$ pixel should be obtainable in an ideal system. However, analysis on the acquired images showed very few dispersion magnitudes below 1-2~arcsec and a general deviation from the expected trend at lower zenith distances. We conclude that it is easy to create PSF elongation, but difficult to recover its origin, without an extensive understanding of the complete system and observing conditions. 

The type of diffraction mask proposed in this work can thus be a valuable tool for the measurement of chromatic dispersion from imaging data. Consequently, it may be used for the validation and calibration of atmospheric dispersion correctors on large telescopes. With higher resolution, diffraction limited image quality, strictly defined passbands and rigorous plate scale calibrations, the uncertainties in the measurement of atmospheric dispersion with speckle method will decrease substantially. This then opens the door to verify or improve upon the existing descriptions of atmospheric dispersion.

\section*{Acknowledgements}
We would like to thank Jake Noel-Storr and Dirk van der Geest for their help and expertise before and during the observation with the Gratama telescope. We would also like to acknowledge Ramon Navarro, Eline Tolstoy, Bayu Jayawardhana, Marc Verheijen, Jason Spyromilio and Davide Massari for useful discussions, feedback and support during this research.
Finally, we would like to thank the anonymous reviewer, whose comments improved several aspects of this paper.

\section*{Data Availability}
We support open science and open data policies and have therefore made all our data and analysis code publicly available on the GitLab page of the Kapteyn Astronomical Institute. The repository includes detailed walkthroughs of the data reduction and main algorithms. It can be found at  \url{https://gitlab.astro.rug.nl/born/ProjectNewton}.



\bibliographystyle{mnras}
\bibliography{references} 

\begin{thebibliography}{}
\makeatletter
\relax
\def\mn@urlcharsother{\let\do\@makeother \do\$\do\&\do\#\do\^\do\_\do\%\do\~}
\def\mn@doi{\begingroup\mn@urlcharsother \@ifnextchar [ {\mn@doi@}
  {\mn@doi@[]}}
\def\mn@doi@[#1]#2{\def\@tempa{#1}\ifx\@tempa\@empty \href
  {http://dx.doi.org/#2} {doi:#2}\else \href {http://dx.doi.org/#2} {#1}\fi
  \endgroup}
\def\mn@eprint#1#2{\mn@eprint@#1:#2::\@nil}
\def\mn@eprint@arXiv#1{\href {http://arxiv.org/abs/#1} {{\tt arXiv:#1}}}
\def\mn@eprint@dblp#1{\href {http://dblp.uni-trier.de/rec/bibtex/#1.xml}
  {dblp:#1}}
\def\mn@eprint@#1:#2:#3:#4\@nil{\def\@tempa {#1}\def\@tempb {#2}\def\@tempc
  {#3}\ifx \@tempc \@empty \let \@tempc \@tempb \let \@tempb \@tempa \fi \ifx
  \@tempb \@empty \def\@tempb {arXiv}\fi \@ifundefined
  {mn@eprint@\@tempb}{\@tempb:\@tempc}{\expandafter \expandafter \csname
  mn@eprint@\@tempb\endcsname \expandafter{\@tempc}}}

\bibitem[\protect\citeauthoryear{{Arribas}, {Mediavilla},
  {Garc{\'\i}a-Lorenzo}, {del Burgo}  \& {Fuensalida}}{{Arribas}
  et~al.}{1999}]{Arribas1999}
{Arribas} S.,  {Mediavilla} E.,  {Garc{\'\i}a-Lorenzo} B.,  {del Burgo} C.,
  {Fuensalida} J.~J.,  1999, \mn@doi [A\&AS] {10.1051/aas:1999463}, \href
  {https://ui.adsabs.harvard.edu/abs/1999A&AS..136..189A} {136, 189}

\bibitem[\protect\citeauthoryear{{Avila}, {Rupprecht}  \& {Beckers}}{{Avila}
  et~al.}{1997}]{Avila1997}
{Avila} G.,  {Rupprecht} G.,   {Beckers} J.~M.,  1997, in {Ardeberg} A.~L.,
  ed.,  Society of Photo-Optical Instrumentation Engineers (SPIE) Conference
  Series Vol. 2871, Optical Telescopes of Today and Tomorrow. pp 1135--1143,
  \mn@doi{10.1117/12.269000}

\bibitem[\protect\citeauthoryear{{Bos}}{{Bos}}{2020}]{Bos2020}
{Bos} S.~P.,  2020, \mn@doi [A\&A] {10.1051/0004-6361/202037957}, \href
  {https://ui.adsabs.harvard.edu/abs/2020A&A...638A.118B} {638, A118}

\bibitem[\protect\citeauthoryear{{Cabral} \& {Wehbe}}{{Cabral} \&
  {Wehbe}}{2021}]{Cabral2020}
{Cabral} A.,  {Wehbe} B.,  2021, \mn@doi [JATIS] {10.1117/1.JATIS.7.3.035003},
  \href {https://ui.adsabs.harvard.edu/abs/2021JATIS...7c5003C} {7, 035003}

\bibitem[\protect\citeauthoryear{{Craig} \& {Chambers}}{{Craig} \&
  {Chambers}}{2021}]{CCDDataReductionGuide}
{Craig} M.,  {Chambers} L.,  2021, {CCD Data Reduction Guide},
  \url{http://www.astropy.org/ccd-reduction-and-photometry-guide/}

\bibitem[\protect\citeauthoryear{{Egner} et~al.,}{{Egner}
  et~al.}{2010}]{Egner2010}
{Egner} S.,  et~al., 2010, in Adaptive Optics Systems II. p. 77364V,
  \mn@doi{10.1117/12.856579}

\bibitem[\protect\citeauthoryear{{F{\'e}tick} et~al.,}{{F{\'e}tick}
  et~al.}{2019}]{Fetick2019}
{F{\'e}tick} R.~J.~L.,  et~al., 2019, \mn@doi [\aap]
  {10.1051/0004-6361/201935830}, \href
  {https://ui.adsabs.harvard.edu/abs/2019A&A...628A..99F} {628, A99}

\bibitem[\protect\citeauthoryear{{Gilles}, {Wang}  \& {Boyer}}{{Gilles}
  et~al.}{2018}]{Gilles2018}
{Gilles} L.,  {Wang} L.,   {Boyer} C.,  2018, in {Close} L.~M.,  {Schreiber}
  L.,   {Schmidt} D.,  eds,  Society of Photo-Optical Instrumentation Engineers
  (SPIE) Conference Series Vol. 10703, Adaptive Optics Systems VI. p. 1070349,
  \mn@doi{10.1117/12.2315793}

\bibitem[\protect\citeauthoryear{Goodman}{Goodman}{2005}]{Goodman2005}
Goodman J.,  2005, Introduction to Fourier Optics.
McGraw-Hill physical and quantum electronics series, W. H. Freeman

\bibitem[\protect\citeauthoryear{Hal{\i}r \& Flusser}{Hal{\i}r \&
  Flusser}{1998}]{halir1998}
Hal{\i}r R.,  Flusser J.,  1998, in Proc. 6th International Conference in
  Central Europe on Computer Graphics and Visualization. WSCG. pp 125--132

\bibitem[\protect\citeauthoryear{{Janssen}}{{Janssen}}{2009}]{Janssen2009}
{Janssen} A.~W.,  2009, Bsc thesis, Kapteyn Astronomical Institute, University
  of Groningen

\bibitem[\protect\citeauthoryear{{Jolissaint}, {Ragland}, {Christou}  \&
  {Wizinowich}}{{Jolissaint} et~al.}{2018}]{Jolissaint2018}
{Jolissaint} L.,  {Ragland} S.,  {Christou} J.,   {Wizinowich} P.,  2018,
  \mn@doi [ApOpt] {10.1364/AO.57.007837}, \href
  {https://ui.adsabs.harvard.edu/abs/2018ApOpt..57.7837J} {57, 7837}

\bibitem[\protect\citeauthoryear{{Jovanovic}, {Guyon}, {Martinache}, {Pathak},
  {Hagelberg}  \& {Kudo}}{{Jovanovic} et~al.}{2015}]{Jovanovic2015}
{Jovanovic} N.,  {Guyon} O.,  {Martinache} F.,  {Pathak} P.,  {Hagelberg} J.,
  {Kudo} T.,  2015, \mn@doi [ApJL] {10.1088/2041-8205/813/2/L24}, \href
  {https://ui.adsabs.harvard.edu/abs/2015ApJ...813L..24J} {813, L24}

\bibitem[\protect\citeauthoryear{{Landolt}}{{Landolt}}{2013}]{Landolt2013}
{Landolt} A.~U.,  2013, \mn@doi [\aj] {10.1088/0004-6256/146/5/131}, \href
  {https://ui.adsabs.harvard.edu/abs/2013AJ....146..131L} {146, 131}

\bibitem[\protect\citeauthoryear{{Lang}, {Hogg}, {Mierle}, {Blanton}  \&
  {Roweis}}{{Lang} et~al.}{2010}]{Lang2010}
{Lang} D.,  {Hogg} D.~W.,  {Mierle} K.,  {Blanton} M.,   {Roweis} S.,  2010,
  \mn@doi [\aj] {10.1088/0004-6256/139/5/1782}, \href
  {https://ui.adsabs.harvard.edu/abs/2010AJ....139.1782L} {139, 1782}

\bibitem[\protect\citeauthoryear{{Le F{\`e}vre} et~al.,}{{Le F{\`e}vre}
  et~al.}{2003}]{VIMOS2003}
{Le F{\`e}vre} O.,  et~al., 2003, in {Iye} M.,  {Moorwood} A. F.~M.,  eds,
  Society of Photo-Optical Instrumentation Engineers (SPIE) Conference Series
  Vol. 4841, Instrument Design and Performance for Optical/Infrared
  Ground-based Telescopes. pp 1670--1681, \mn@doi{10.1117/12.460959}

\bibitem[\protect\citeauthoryear{{Martinez}, {Kolb}, {Tokovinin}  \&
  {Sarazin}}{{Martinez} et~al.}{2010}]{Martinez2010}
{Martinez} P.,  {Kolb} J.,  {Tokovinin} A.,   {Sarazin} M.,  2010, \mn@doi
  [A\&A] {10.1051/0004-6361/201014413}, \href
  {https://ui.adsabs.harvard.edu/abs/2010A&A...516A..90M} {516, A90}

\bibitem[\protect\citeauthoryear{{Massari}, {Marasco}, {Beltramo-Martin},
  {Milli}, {Fiorentino}, {Tolstoy}  \& {Kerber}}{{Massari}
  et~al.}{2020}]{Massari2020}
{Massari} D.,  {Marasco} A.,  {Beltramo-Martin} O.,  {Milli} J.,  {Fiorentino}
  G.,  {Tolstoy} E.,   {Kerber} F.,  2020, \mn@doi [A\&A]
  {10.1051/0004-6361/201937359}, \href
  {https://ui.adsabs.harvard.edu/abs/2020A&A...634L...5M} {634, L5}

\bibitem[\protect\citeauthoryear{{Moffat}}{{Moffat}}{1969}]{Moffat1969}
{Moffat} A.~F.~J.,  1969, A\&A, \href
  {https://ui.adsabs.harvard.edu/abs/1969A&A.....3..455M} {3, 455}

\bibitem[\protect\citeauthoryear{{Pathak} et~al.,}{{Pathak}
  et~al.}{2016}]{Pathak2016}
{Pathak} P.,  et~al., 2016, \mn@doi [\pasp] {10.1088/1538-3873/128/970/124404},
  \href {https://ui.adsabs.harvard.edu/abs/2016PASP..128l4404P} {128, 124404}

\bibitem[\protect\citeauthoryear{{Pathak} et~al.,}{{Pathak}
  et~al.}{2018}]{Pathak2018}
{Pathak} P.,  et~al., 2018, \mn@doi [\pasp] {10.1088/1538-3873/aa96f9}, \href
  {https://ui.adsabs.harvard.edu/abs/2018PASP..130b5004P} {130, 025004}

\bibitem[\protect\citeauthoryear{{Perryman} et~al.,}{{Perryman}
  et~al.}{1997}]{Hipparcos1997}
{Perryman} M.~A.~C.,  et~al., 1997, A\&A, \href
  {https://ui.adsabs.harvard.edu/abs/1997A&A...323L..49P} {500, 501}

\bibitem[\protect\citeauthoryear{{Phillips}, {Suzuki}, {Larkin}, {Moore},
  {Hayano}, {Tsuzuki}  \& {Wright}}{{Phillips} et~al.}{2016}]{Phillips2016}
{Phillips} A.~C.,  {Suzuki} R.,  {Larkin} J.~E.,  {Moore} A.~M.,  {Hayano} Y.,
  {Tsuzuki} T.,   {Wright} S.~A.,  2016, in {Evans} C.~J.,  {Simard} L.,
  {Takami} H.,  eds,  Society of Photo-Optical Instrumentation Engineers (SPIE)
  Conference Series Vol. 9908, Ground-based and Airborne Instrumentation for
  Astronomy VI. p. 9908A1 (\mn@eprint {arXiv} {1608.01690}),
  \mn@doi{10.1117/12.2232952}

\bibitem[\protect\citeauthoryear{{Simioni} et~al.,}{{Simioni}
  et~al.}{2020}]{Simioni2020}
{Simioni} M.,  et~al., 2020, in Society of Photo-Optical Instrumentation
  Engineers (SPIE) Conference Series. p. 1144837 (\mn@eprint {arXiv}
  {2012.04959}), \mn@doi{10.1117/12.2561251}

\bibitem[\protect\citeauthoryear{{Skemer} et~al.,}{{Skemer}
  et~al.}{2009}]{Skemer2009}
{Skemer} A.~J.,  et~al., 2009, \mn@doi [\pasp] {10.1086/605312}, \href
  {https://ui.adsabs.harvard.edu/abs/2009PASP..121..897S} {121, 897}

\bibitem[\protect\citeauthoryear{{Spyromilio}}{{Spyromilio}}{2021}]{spyromilio2021}
{Spyromilio} J.,  2021, Private communication

\bibitem[\protect\citeauthoryear{{Traa}}{{Traa}}{2013}]{Traa2013}
{Traa} J.,  2013, {Least-Squares Intersection of Lines},
  \url{https://silo.tips/download/least-squares-intersection-of-lines}

\bibitem[\protect\citeauthoryear{\VAN{Born}{Van den}{van den}~Born \&
  {Jellema}}{\VAN{Born}{Van den}{van den}~Born \&
  {Jellema}}{2020}]{vandenBorn2020}
\VAN{Born}{Van den}{van den}~Born J.~A.,  {Jellema} W.,  2020, \mn@doi [\mnras]
  {10.1093/mnras/staa1870}, \href
  {https://ui.adsabs.harvard.edu/abs/2020MNRAS.496.4266V} {496, 4266}

\bibitem[\protect\citeauthoryear{\VAN{Horst}{Ter Horst}{ter Horst}, {Kragt},
  {Lesman}  \& {Navarro}}{\VAN{Horst}{Ter Horst}{ter Horst}
  et~al.}{2016}]{terHorst2016}
\VAN{Horst}{Ter Horst}{ter Horst} R.,  {Kragt} J.,  {Lesman} D.,   {Navarro}
  R.,  2016, in {Navarro} R.,  {Burge} J.~H.,  eds,  Society of Photo-Optical
  Instrumentation Engineers (SPIE) Conference Series Vol. 9912, Advances in
  Optical and Mechanical Technologies for Telescopes and Instrumentation II. p.
  99121J, \mn@doi{10.1117/12.2232348}

\bibitem[\protect\citeauthoryear{{Wang} et~al.,}{{Wang}
  et~al.}{2014}]{Wang2014}
{Wang} J.~J.,  et~al., 2014, in {Ramsay} S.~K.,  {McLean} I.~S.,   {Takami} H.,
   eds,  Society of Photo-Optical Instrumentation Engineers (SPIE) Conference
  Series Vol. 9147, Ground-based and Airborne Instrumentation for Astronomy V.
  p. 914755 (\mn@eprint {arXiv} {1407.2308}), \mn@doi{10.1117/12.2055753}

\bibitem[\protect\citeauthoryear{{Wehbe}, {Cabral}  \& {{\'A}vila}}{{Wehbe}
  et~al.}{2020}]{Wehbe2020}
{Wehbe} B.,  {Cabral} A.,   {{\'A}vila} G.,  2020, \mn@doi [\mnras]
  {10.1093/mnras/staa2726}, \href
  {https://ui.adsabs.harvard.edu/abs/2020MNRAS.499..183W} {499, 183}

\bibitem[\protect\citeauthoryear{{Wenger} et~al.,}{{Wenger}
  et~al.}{2000}]{SIMBAD2000}
{Wenger} M.,  et~al., 2000, \mn@doi [A\&AS] {10.1051/aas:2000332}, \href
  {https://ui.adsabs.harvard.edu/abs/2000A&AS..143....9W} {143, 9}

\bibitem[\protect\citeauthoryear{{Wilby}, {Keller}, {Snik}, {Korkiakoski}  \&
  {Pietrow}}{{Wilby} et~al.}{2017}]{Wilby2017}
{Wilby} M.~J.,  {Keller} C.~U.,  {Snik} F.,  {Korkiakoski} V.,   {Pietrow}
  A.~G.~M.,  2017, \mn@doi [A\&A] {10.1051/0004-6361/201628628}, \href
  {https://ui.adsabs.harvard.edu/abs/2017A&A...597A.112W} {597, A112}

\bibitem[\protect\citeauthoryear{{Zandvliet}}{{Zandvliet}}{2017}]{Zandvliet2017}
{Zandvliet} M.,  2017, Msc thesis, Kapteyn Astronomical Institute, University
  of Groningen

\makeatother
\end{thebibliography}



\appendix



\section{Observation details}\label{appendix:observations}
\begin{figure*}
\centering
\begin{tabular}{ccc}
\includegraphics[width=0.3\textwidth]{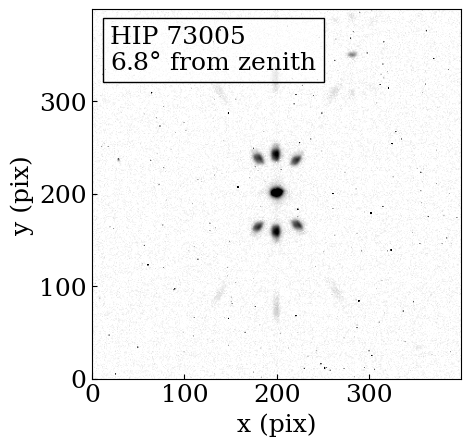} & 
\includegraphics[width=0.3\textwidth]{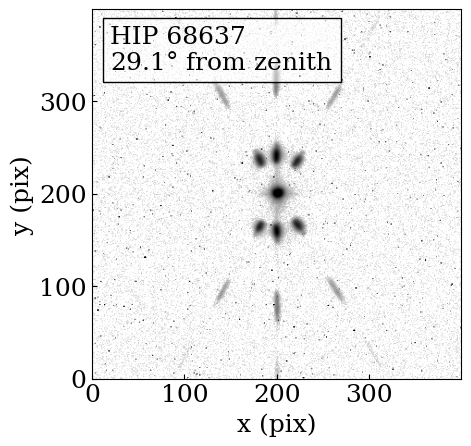} & \includegraphics[width=0.3\textwidth]{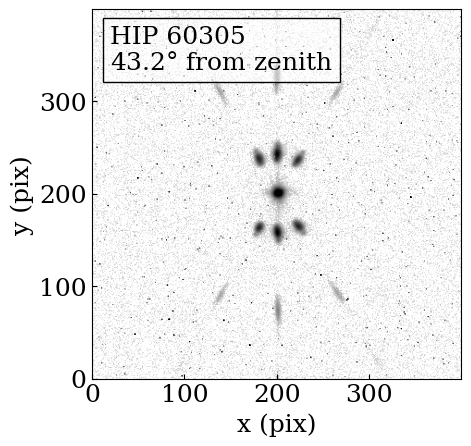} \\

\includegraphics[width=0.3\textwidth]{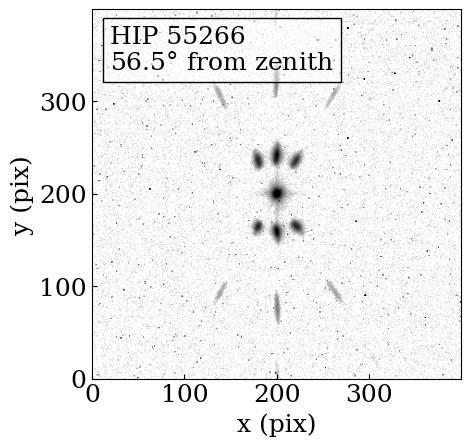} & 
\includegraphics[width=0.3\textwidth]{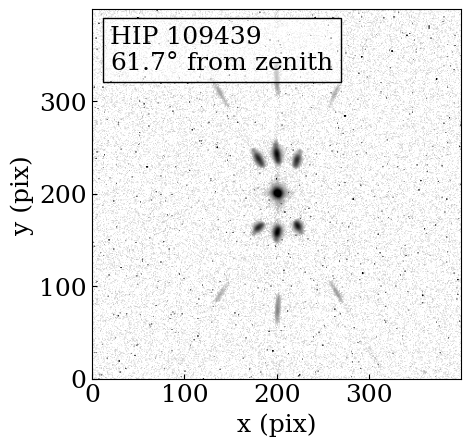} & \includegraphics[width=0.3\textwidth]{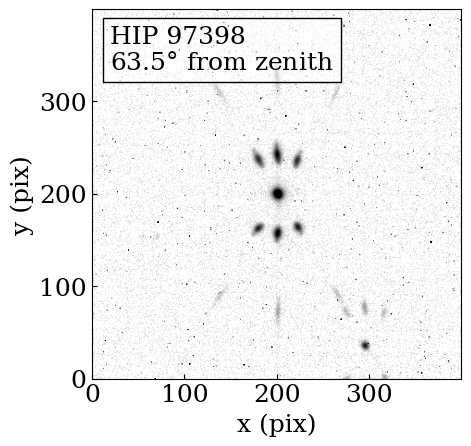} \\

\includegraphics[width=0.3\textwidth]{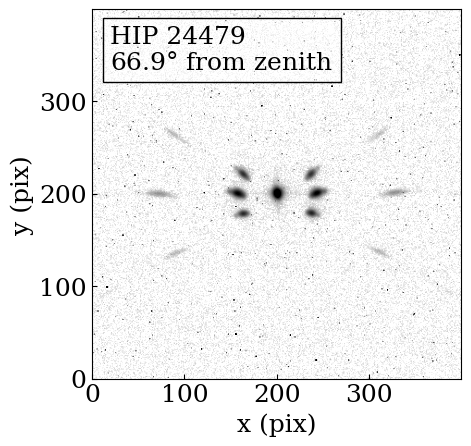} & 
\includegraphics[width=0.3\textwidth]{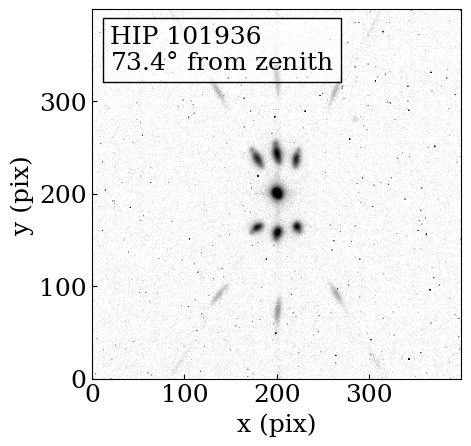} & \includegraphics[width=0.3\textwidth]{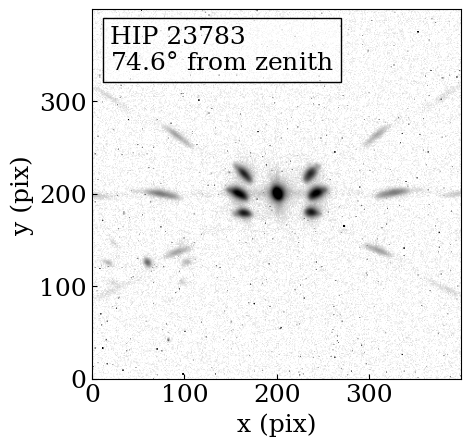} \\
\end{tabular}
\caption{\textit{B}-band images of the nine stars observed on the night of 2021 June 13, sorted by zenith distance. Note that the mask was reoriented during the night. This explains the different orientation of the speckles in two of the nine images.}\label{fig:appendix_B-band_images}
\end{figure*}

Table~\ref{tab:observation_properties} provides an overview of all the targets that were observed during the two nights at the Blaauw observatory. Figure~\ref{fig:appendix_B-band_images} shows the nine stars observed during the first night in \textit{B}-band. These images illustrate how the location of the radiation centre changes as the zenith distance increases, but also that multiple diffraction orders can easily be distinguished.

\begin{table*}
\caption{Detailed observation log. The major difference in the integration times between \textit{B}- and \textit{V}-band can be explained by the decreased response of the telescope and CCD to the shorter wavelengths in \textit{B}-band. The \textit{B} and \textit{V} magnitudes are given as presented by the SIMBAD database \citep{SIMBAD2000}, except for HIP 23783, which is a variable star with \textit{V}$\approx 4.9-5$.} The zenith distance is given for the start of the observation of each object.
\label{tab:observation_properties}
\begin{tabular}{l l l r r r r r r r}
\hline
Date \& Time (UTC) & Object & Type & \textit{B} & \textit{V} & RA (ICRS) & DEC (ICRS) & Zenith distance & $t_{\textnormal{exp}}(B)$ & $t_{\textnormal{exp}}(V)$\\
\hline
2021 June 13 - 21:47    & HIP \phantom{1}73005 & K1V      & 8.59  & 7.80  & 14h 55m 11.044s & 53\degr 40\arcmin 49.247\arcsec &  6.9\degr & 3x40s & 3x30s \\
2021 June 13 - 22:18    & HIP \phantom{1}97398 & K0       & 7.79  & 6.72  & 19h 47m 46.149s &  5\degr 46\arcmin 55.185\arcsec & 64.3\degr & 3x20s & 3x12s \\
2021 June 13 - 22:35    & HIP 101936           & K0III    & 6.20  & 5.15  & 20h 39m 24.893s &  0\degr 29\arcmin 11.133\arcsec & 73.4\degr & 3x20s & 3x5s  \\
2021 June 13 - 22:51    & HIP \phantom{1}55266 & A2V      & 4.90  & 4.80  & 11h 19m  7.890s & 38\degr 11\arcmin 08.046\arcsec & 55.8\degr & 3x4s  & 3x3s \\
2021 June 13 - 23:05    & HIP \phantom{1}60305 & K1III    & 7.45  & 6.33  & 12h 21m 56.250s & 47\degr 10\arcmin 55.931\arcsec & 43.2\degr & 3x20s & 3x10s \\
2021 June 13 - 23:15    & HIP \phantom{1}68637 & A2V      & 6.16  & 6.15  &  4h 02m 59.758s & 50\degr 58\arcmin 18.510\arcsec & 28.4\degr & 3x13s & 3x10s \\
2021 June 13 - 23:40    & HIP 109439           & G1V      & 6.87  & 6.18  & 22h 10m 19.022s & 19\degr 36\arcmin 58.821\arcsec & 61.7\degr & 3x15s & 3x9s \\
2021 June 14 - 00:04    & HIP \phantom{1}23783 & F2V      &       &       &  5h 06m 40.631s & 51\degr 35\arcmin 51.805\arcsec & 74.7\degr & 3x20s & 3x7s \\
2021 June 14 - 00:08    & HIP \phantom{1}24479 & K2III-IV & 7.33  & 6.14  &  5h 15m 11.402s & 59\degr 24\arcmin 20.477\arcsec & 66.9\degr & 3x20s & 3x12s \\
2021 October 24 - 18:27 & HIP 105867           & B9       & 7.58  & 7.60  & 21h 26m 29.089s & 52\degr 44\arcmin 52.641\arcsec &  3.1\degr & 3x40s & 3x30s \\
2021 October 24 - 18:47 & HIP 110924           & B9       & 7.73  & 7.73  & 22h 28m 25.537s & 51\degr 18\arcmin 22.447\arcsec &  9.6\degr & 3x70s & 3x60s \\
2021 October 24 - 19:11 & HIP \phantom{1}99506 & B9       & 7.74  & 7.71  & 20h 11m 39.357s & 63\degr 31\arcmin 38.692\arcsec & 16.4\degr & 3x70s & 3x40s \\
2021 October 24 - 19:31 & HIP \phantom{12}1925 & B9       & 7.73  & 7.58  & 00h 24m 17.453s & 60\degr 27\arcmin 11.021\arcsec & 19.4\degr & 3x60s & 3x45s \\
2021 October 24 - 19:52 & HIP 114178           & B9       & 7.85  & 7.90  & 23h 07m 20.411s & 80\degr 38\arcmin 15.231\arcsec & 27.7\degr & 3x60s & 3x45s \\
2021 October 24 - 20:11 & HIP \phantom{1}19377 & B9       & 7.64  & 7.61  & 04h 09m 03.109s & 71\degr 20\arcmin 37.259\arcsec & 37.6\degr & 3x60s & 3x45s \\
2021 October 24 - 20:26 & HIP \phantom{1}25328 & B9       & 8.03  & 7.96  & 05h 25m 01.257s & 61\degr 49\arcmin 32.284\arcsec & 47.1\degr & 3x60s & 3x45s \\
2021 October 24 - 20:41 & HIP \phantom{1}34825 & B9       & 7.57  & 7.58  & 07h 12m 27.977s & 62\degr 15\arcmin 44.842\arcsec & 54.7\degr & 3x60s & 3x45s \\
2021 October 24 - 21:12 & HIP \phantom{1}29332 & B9       & 7.76  & 7.65  & 06h 11m 03.660s & 41\degr 13\arcmin 56.987\arcsec & 60.7\degr & 3x60s & 3x45s \\
2021 October 24 - 21:29 & HIP \phantom{1}32757 & B9       & 7.68  & 7.75  & 06h 49m 49.993s & 39\degr 28\arcmin 44.768\arcsec & 64.8\degr & 3x60s & 3x45s \\
2021 October 24 - 21:49 & HIP \phantom{1}36066 & B9       & 7.65  & 7.54  & 07h 25m 55.702s & 32\degr 51\arcmin 39.999\arcsec & 72.0\degr & 3x60s & 3x45s \\
\hline
\end{tabular}
\end{table*}

\section{Derivation of the Least Squares Intersection of multiple lines}\label{appendix:least_squares_intersection}
In this appendix we describe the derivation of the least squares distance of many lines to a common intersection point. This derivation is described by \cite{Traa2013}, but was not yet available in published literature known to the authors.

\citeauthor{Traa2013} starts with the definition of a two-dimensional line. In vector notation a point $\boldsymbol{a}$ on the line,
\begin{equation}
    \boldsymbol{a} = \left[a_1, a_2\right]^T,
\end{equation}
and a direction vector $\boldsymbol{n}$,
\begin{align}
    \boldsymbol{n} &= \left[n_1, n_2 \right]^T, &&\textnormal{with } ||\boldsymbol{n}||_2 = \boldsymbol{n}^T\boldsymbol{n} = 1
\end{align}
define the line
\begin{align}
    \boldsymbol{p} &= \boldsymbol{a} + t \boldsymbol{n},  &&\textnormal{where } -\infty < t < \infty. 
\end{align}

The squared perpendicular distance of a point $\boldsymbol{p}$ to a line defined by $\boldsymbol{a}$ and $\boldsymbol{n}$ is found from
\begin{align}
    D(\boldsymbol{p}; \boldsymbol{a}, \boldsymbol{n} ) &= ||(\boldsymbol{a} - \boldsymbol{p}) - ((\boldsymbol{a} - \boldsymbol{p})^T\boldsymbol{n})\boldsymbol{n}||_2^2 \nonumber\\
    &= || (\boldsymbol{a} - \boldsymbol{p}) - \boldsymbol{n}\boldsymbol{n}^T (\boldsymbol{a} - \boldsymbol{p}) ||_2^2 \nonumber\\
    &= || (\boldsymbol{I} - \boldsymbol{n}\boldsymbol{n}^T) (\boldsymbol{a} - \boldsymbol{p})||_2^2 \nonumber\\
    &= (\boldsymbol{a} - \boldsymbol{p})^T (\boldsymbol{I} - \boldsymbol{n}\boldsymbol{n}^T) (\boldsymbol{a} - \boldsymbol{p}).
\end{align}
The matrix $\boldsymbol{I} - \boldsymbol{n}\boldsymbol{n}^T$ projects the vectors $\boldsymbol{a}$ and $\boldsymbol{p}$ into a space orthogonal to $\boldsymbol{n}$. \citeauthor{Traa2013} then equates minimising the distance to the line in the projected space to a maximisation of the pdf of a Gaussian distribution. In that case a least squares solution must be possible.

To find the best intersection point for $K$ lines, \citeauthor{Traa2013} minimises the sum of the squared distances including a confidence level $c_j$ for each line $j$.
\begin{align}
    D(\boldsymbol{p}; \boldsymbol{A}, \boldsymbol{N}, \boldsymbol{c}) &= \sum^K_{j=1} D(\boldsymbol{p}; \boldsymbol{a}_j, \boldsymbol{n}_j, c_j) \nonumber\\
    &= \sum^K_{j=1}c_j (\boldsymbol{a}_j - \boldsymbol{p})^T (\boldsymbol{I} - \boldsymbol{n}_j \boldsymbol{n}_j^T) (\boldsymbol{a}_j - \boldsymbol{p}).
\end{align}
Now, the objective is to find a point where this sum of squared distances is minimised. Thus
\begin{equation}
    \hat{\boldsymbol{p}} = \argmin_{\boldsymbol{p}} D (\boldsymbol{p}; \boldsymbol{A}, \boldsymbol{N}, \boldsymbol{c}).
\end{equation}

This problem is quadratic in $\boldsymbol{p}$ and can therefore be solved by finding the point at which the derivative becomes zero.
\begin{equation}
    \frac{\partial D}{\partial \boldsymbol{p}} = \sum^K_{j=1} -2 c_j (\boldsymbol{I} - \boldsymbol{n}_j \boldsymbol{n}_j^T ) (\boldsymbol{a}_j - \boldsymbol{p}) = 0
\end{equation}
Rearranging this expression leads to a linear set of equations of the form
\begin{align}
    \boldsymbol{R} \boldsymbol{p} = \boldsymbol{q},
\end{align}
where
\begin{align}
    \boldsymbol{R} = \sum^K_{j=1} c_j(\boldsymbol{I} - \boldsymbol{n}_j\boldsymbol{n}_j^T), && \boldsymbol{q} = \sum^K_{j=1} c_j (\boldsymbol{I} - \boldsymbol{n}_j \boldsymbol{n}_j^T) \boldsymbol{a}_j.
\end{align}
We may solve this system directly or with the help of a Moore-Penrose pseudoinverse of $\boldsymbol{R}$. This finally provides us the location of the (weighted) best intersection point,
\begin{equation}
    \hat{\boldsymbol{p}} = \boldsymbol{R}^{\dagger} \boldsymbol{q}.
\end{equation}

\bsp	
\label{lastpage}
\end{document}